\documentclass[a4paper,11pt]{article}
\usepackage{jcappub} % for details on the use of the package, please see the JINST-author-manual
\usepackage{graphicx}
\usepackage{amssymb,amsmath}
\usepackage{subfiles}
\usepackage{hyperref}
\usepackage{footmisc}
\usepackage{booktabs}
\usepackage{lineno}
\usepackage[utf8]{inputenc}
\usepackage{graphicx}
\usepackage{xcolor,bm}
\usepackage[T1]{fontenc}
\usepackage[english]{babel}
\usepackage[utf8]{inputenc}
\usepackage[english]{babel}
\DeclareUnicodeCharacter{2212}{-}

\usepackage[letterpaper,top=2cm,bottom=2cm,left=3cm,right=3cm,marginparwidth=1.75cm]{geometry}
\title{The complementary of CTAO, direct detection and collider searches for dark matter in Effective Field Theories and Simplified models}
\author[a,b]{Igor Reis}
\author[c]{Andre Scaffidi}
\author[b]{Emmanuel Moulin}
\author[d]{Martin White}

\affiliation[a]{Universidade de São Paulo, Instituto de Física de São Carlos, Av. Trabalhador São Carlense 400, São Carlos, Brazil}
\affiliation[b]{Irfu, CEA Saclay, Université Paris-Saclay, F-91191 Gif-sur-Yvette, France}
\affiliation[c]{Theoretical and Scientific Data Science, Scuola Internazionale Superiore di Studi Avanzati (SISSA), 34136 Trieste, Italy}
\affiliation[d]{ARC Centre of Excellence for Dark Matter Particle Physics \& CSSM,
Department of Physics, University of Adelaide, Adelaide, SA 5005, Australia}

\emailAdd{igorreis@ifsc.usp.br}
\emailAdd{igor.reis@cea.fr}
\emailAdd{ascaffid@sissa.it}
\emailAdd{emmanuel.moulin@cea.fr}
\emailAdd{martin.white@adelaide.edu.au}

\abstract{This paper explores the sensitivity of the Cherenkov Telescope Array Observatory to dark matter annihilation in the Galactic Center, within the frameworks of Effective Field Theory and Simplified Models. We present sensitivity forecasts, utilizing an up-to-date instrument configuration and incorporating the latest models for Galactic Diffuse Emission. A key aspect of our work is the inclusion of updated dark matter density profiles, J-factors, and velocity dispersion distributions derived from the FIRE-2 cosmological hydrodynamical simulations, which significantly impact the expected indirect detection signals. Furthermore, we update the constraints from direct detection experiments (Xenon1T and LZ)  taking into account the astrophysical uncertainties informed by the FIRE-2 simulations, and also investigate limits coming from collider searches (ATLAS and CMS). Our analysis reveals improved constraints on the effective suppression scale ($M_*$) in the Effective Field Theory framework and on the mediator mass ($M_{med}$) in Simplified Models compared to previous studies, highlighting the complementarity of the Cherenkov Telescope Array Observatory with direct and collider searches in probing a wide range of dark matter scenarios. We discuss the implications of these results for various dark matter interaction types, including scalar, pseudoscalar, vector, and axial-vector mediators, and emphasize the importance of considering realistic astrophysical inputs in interpreting dark matter search results across different experimental fronts.}

\begin{document}

\maketitle

\newpage

\section{Introduction}
%%%%%%%%%%%%%%%%%%%%
%%%%%%%%%%%%%%%%%%%%
%%%%%%%%%%%%%%%%%%%%
Astrophysical and cosmological measurements strongly support the view that 85\% of the total matter content of the universe is comprised of an elusive form of ``dark matter'' (DM). However, little is known about its nature and its interactions with Standard Model (SM) particles. 
Among the most compelling hypotheses for DM is that it results from a new type of Weakly Interacting Massive Particle (WIMP), with weak-scale interactions and a mass
in the GeV-TeV range. A simple thermal production mechanism allows such a WIMP to be produced with an abundance that matches the cosmological observation~\cite{Lee:1977ua}, often referred to as the ``WIMP miracle''.

Huge experimental efforts have been deployed over several decades to probe WIMP interactions with Standard Model species: i) Collider searches look for the direct production of WIMPs in final states with missing energy, plus evidence of a mediator for DM-SM interactions~\cite{cms2012search,collaboration2013search,aaltonen2012search,ATLAS:2024bms,ATLAS:2024npu,ATLAS:2024rlu,ATLAS:2024xne,ATLAS:2022znu,ATLAS:2022bzt,ATLAS:2022ygn,ATLAS:2021gcn,ATLAS:2021pdg,ATLAS:2021jbf,ATLAS:2021kxv,ATLAS:2021hza,ATLAS:2020yzc,ATLAS:2020uiq,ATLAS:2020fgc,ATLAS:2019cid,CMS:2024zqs,CMS:2020ulv,CMS:2021far}. ii) Direct searches look for scattering off nuclei or electrons in underground detectors. These experiments utilize different target materials, such as liquid xenon—employed by experiments like XENONnT, LUX-ZEPLIN (LZ), and PandaX, or solid-state detectors, as used in CDMS, CoGeNT~\cite{XENON:2018voc,XENON:2024xgd,aalbers2024dark,pandax2018,cdms2015,cogent2013,dama2018}. iii) Indirect searches aim to detect the final-state products of WIMP annihilation or decay such as gamma rays, neutrinos and charged particles~\cite{HESS:2018pbp,HESS:2022ygk,ANTARES:2019svn,IceCube:2023ies,Xu:2020zfh,DAMPE:2021hsz}.
Exploiting the synergy between these techniques is expected to improve the discovery potential, and would also permit a deeper understanding of the detailed particle physics and associated astrophysical uncertainties~\cite{Baltz:2006fm,White:2010jp}. Whilst individual experiments often present results on phenomenological quantities that are directly constrained by the relevant data (\textit{e.g.}, the WIMP mass and the annihilation or scattering rate), a direct combination of results from multiple probes requires a concrete particle physics hypothesis. A relatively agnostic approach is to create a ``bottom-up'' DM theory in which a WIMP candidate is added by hand to the SM, along with the necessary interactions with SM fields. The simplest way to do this is to use the familiar framework of effective field theory (EFT)~\cite{PhysRevD.85.056011} in which one assumes that the SM is the lowest order contribution to a theory of new physics which is associated with a mass scale $M_\star$. The SM Lagrangian density can then be supplemented with a series of effective operators built using SM fields, a new dark matter field, and potentially derivatives of fields, each suppressed by inverse powers of $M_\star$ to ensure that each term in the Lagrangian density retains a mass dimension of four. Such a theory becomes invalid if the mass scale $M_\star$ is comparable to the momentum exchange in the experiment under consideration~\cite{Malik_2015,Buchmueller_2015,Harris_2015}. For this reason, and specifically driven by the requirements of interpreting data from high energy collider experiments, a more recent approach has been that of ``simplified models'', in which an explicit mediator particle is added to the SM in addition to the WIMP. Comprehensive statistical analyses of a DM effective theory and simplified models with $s$-channel vector mediators have recently been performed~\cite{GAMBIT:2021rlp,Chang:2022jgo,Chang:2023cki} and a recent review of theory development for the interpretation of DM collider searches is presented in Ref.~\cite{DeRoeck:2024fjq}. 

The next frontier in indirect searches for DM will be reached in the near future through the construction and commissioning of the Cherenkov Telescope Array Observatory (CTAO). For WIMP masses in the TeV range, CTAO is expected to surpass the present constraints on the annihilation cross section by one order of magnitude~\cite{HESS:2022ygk} and to provide a unprecedented synergy with direct detection and collider searches as highlighted, for instance, in the context of supersymmetry models in Refs.~\cite{Cahill-Rowley:2014boa,Hryczuk:2019nql,Ellis:2022emx}. Given our lack of knowledge of detailed WIMP properties, it is crucial to explore the CTAO reach for realistic and broad WIMP hypotheses such as those described by the EFT and simplified model frameworks. An early study of the CTAO reach for such scenarios can be found in Ref.~\cite{Balazs:2017hxh}, but much has happened in the intervening 7 years to make many of the earlier results obsolete. Our present work improves over the previous one in the following aspects:

\begin{itemize}
\item We make use of an up-to-date configuration of the CTAO Southern-site array, namely the \textit{Alpha} configuration, composed of 14 medium-sized telescopes and 37 small-sized telescopes, using the latest Instrument Response Functions to make a realistic sensitivity forecast. In particular, we make use of the most up-to-date cosmic ray
(CR) background coming from a full Monte Carlo simulation for the \textit{Alpha} configuration. 

\item In order to model the Galactic Diffuse Emission  
expected in the TeV energy range, we utilize 
the most recent release of the CR
propagation framework GALPROP (version 57)~\cite{Porter:2021tlr} assuming  realistic CR source density and interstellar radiation field distributions.

\item Our knowledge of the DM spatial density and velocity distributions within the Milky Way, which is the leading source of uncertainty affecting calculations of gamma ray yields, has improved thanks to refinements in cosmological hydrodynamical simulations of Milky Way-like galaxies. In particular, investigations of the role of baryonic feedback in such simulations has led to substantial deviations from the spatial distributions predicted by DM-only simulations in the central kiloparsecs of the Galaxy, whilst leading to a generic expectation of an increased velocity dispersion. We investigate the effect on the CTAO WIMP sensitivity through use of J-factors\footnote{The J-factor is defined as the integral of the square of the DM density along the line of sight and over the solid angle $\Delta\Omega$ of the observed region (see Section \ref{sec:ind_detec}.} from the Fire-2 simulations~\cite{McKeown:2021sob}. Here, we make use of the well-known fact that the dark matter annihilation cross section can be expressed as a partial-wave expansion in powers of the square of the relative velocity between the annihilating particles. The lowest order term (known as the ``$s$-wave'' term) will dominate when it is physically allowed\footnote{Some DM models posses symmetries that impose s-wave annihilation suppression, such as secluded DM and fermionic Higgs portal DM~\cite{Montanari:2023sln}}, since the next term (the ``$p$-wave'' term) is suppressed by the square of the relative velocity $v$, where $v$ is relatively small for cold dark matter annihilating in galaxies. We use the dedicated J-factors for $s$-wave and $p$-wave annihilation, in the process correcting errors in the treatment of $p$-wave annihilation present in the previous literature~\cite{Balazs:2017hxh}. 

\item Improved constraints on the DM spatial and velocity distributions also have a direct impact on interpreting results from underground direct searches. Specifically, the dispersion of the assumed Galactic velocity distribution of the DM will directly effect the expected scattering rates against nuclei in such detectors. 

\item Both collider and direct search experiment limits on WIMPs have both increased significantly since 2017. We compare our new sensitivity predictions of the CTAO reach with updated limits from the leading experiments in these fields, \textit{i.e.}, ATLAS and CMS for the Large Hadron Collider (LHC), and Xenon1T and LZ for direct detection.
\end{itemize}

The paper is organized as follows. In Sec.~\ref{sec:theory} we describe our assumed WIMP theories in detail, covering both EFT and simplified models. We also detail relevant theoretical aspects   of the experiments we incorporate in our analysis. In Sec.~\ref{sec:signals}, we review our treatment of the CTAO, direct detection and LHC constraints that we use in our analysis. We present and discuss our main results in Sec.~\ref{sec:results}, and present final conclusions in Sec.~\ref{sec:conclusions}.

\section{WIMP models and expected signals}
\label{sec:theory}

\subsection{Dark matter distribution in the Milky Way}

The DM signals expected in direct and indirect detection experiments strongly depend on the DM density and velocity distributions. The scattering rate in direct detection depends linearly on the DM density and velocity distribution in the Solar neighbourhood. The annihilation signal in the GC region
depends critically on the distribution of DM around the centre of the Milky Way which is the result of the complex interplay of DM and baryons during the scaffolding of the DM halo over Gyr evolution. At present, it is neither firmly predicted from cosmological N-body simulations of Milky way-sized galaxies nor significantly constrained from mass-modelling approaches using the baryonic mass distribution. 

DM-only simulations motivate density profiles rising steeply towards the GC for which the NFW~\cite{Navarro:1996gj} and Einasto~\cite{Springel:2008by} functions provide satisfying parametrisations. Incorporating baryon feedback in simulations requires modelling effects such as radiative heating, cooling for gas, stellar feedback, radiation pressure, and star formation, which substantially increases their complexity. Recent realistic 
hydrodynamical simulations, including baryonic effects, have revealed significant deviations from the spatial DM distribution predicted by DM-only simulations, demonstrating that these effects can flatten out the DM density distribution in the inner halo, producing a constant-density kpc-sized core~\cite{Chan:2015tna}.
A generic feature of recent hydrodynamical simulations~\cite{McKeown:2021sob,Board:2021bwj,Hussein:2025xwm} is an increase of the DM density in the inner kpcs
of the GC compared to the NFW and Einasto profiles. A DM core would eventually develop depending
on the level of the baryonic feedback. Such a DM density enhancement in the central kpcs is also observed in mass
modelling approaches~\cite{Portail:2016vei,Lin:2019yux,Cautun:2019eaf}.

For direct detection, the expected scattering rate  depends on the scattering cross section between the DM and the target nucleus, the incoming flux of the DM particles, and on the response function of the detector. The incoming  DM flux is determined by the velocity distribution of the DM particles in the rest frame of the detector. A Maxwell-Boltzmann velocity distribution with a mean value of 220 km/s is usually assumed. Recent hydrodynamical simulations of Milky Way-like galaxies, \textit{e.g.} Ref.~\cite{McKeown:2021sob, Board:2021bwj}, showed that the velocity distribution can be well parametrized by a
Maxwell-Boltzmann with 
an increased mean velocity of about 10\% compared to the commonly assumed value in the Solar neighbourhood~\cite{Schumann:2019eaa}.

\begin{figure}
\hspace{-1.5cm}
    \includegraphics[scale = 0.38]{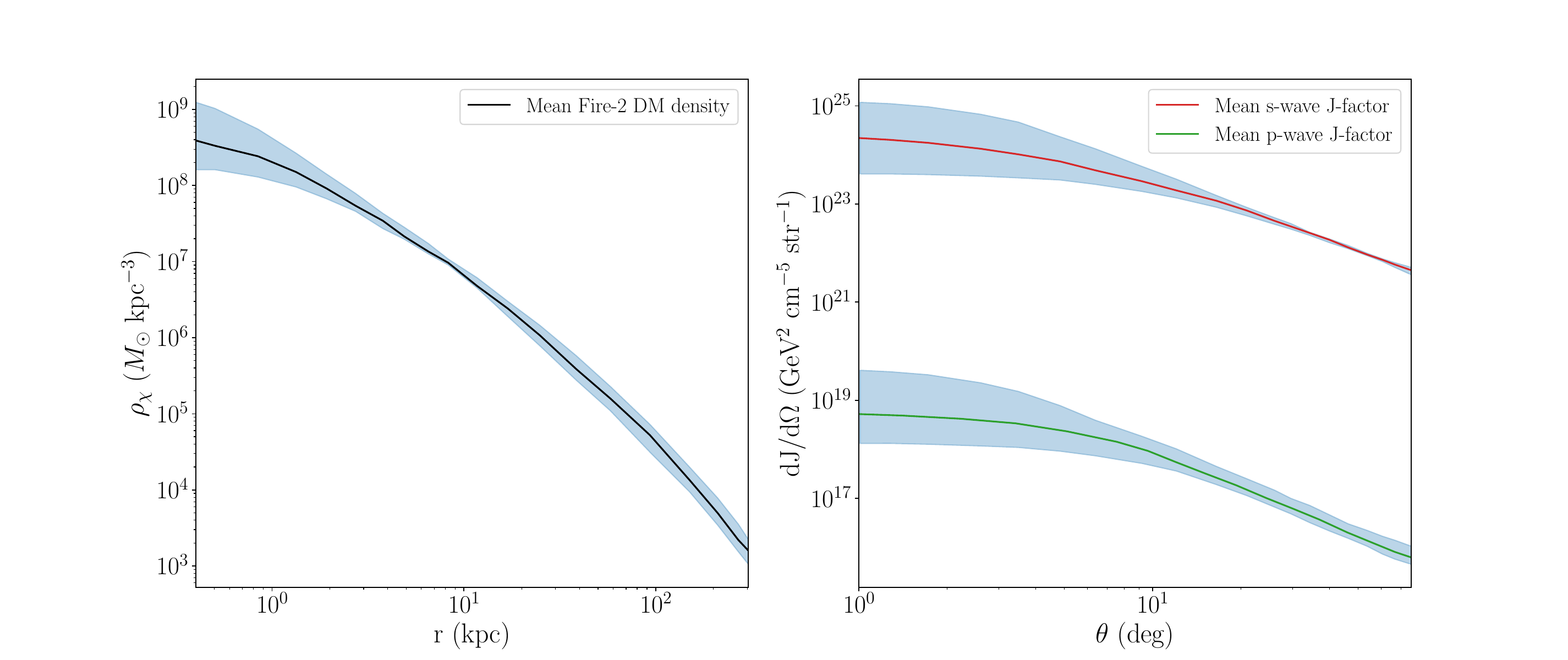}
    \caption{\textit{Left panel}: DM density distribution in the GC region versus distance $r$ from the GC extracted from Fire-2 simulations of Milky Way-like galaxies. The black solid line shows the mean DM density, and the blue shaded region represents the range of possibilities from the set of simulation results of  Ref.~\cite{McKeown:2021sob}. \textit{Right panel}: Corresponding s-wave and p-wave J-factors as a function of the angular distance $\theta$ from the GC. The red and green solid lines stand for the mean J-factor for the s-wave and p-wave cases, respectively. The blue shaded region represents the possible range of values for the J-factors from the same set of simulations. For a reference on the data used in this Figure, see Ref.~\cite{Hopkins:2017ycn}.}
    \label{fig:fire2Jfact}
\end{figure}

Several hydrodynamical simulations have been performed in the context of the FIRE-2 zoom-in simulations~\cite{Hopkins:2017ycn}.  
For each of the 12 pairs of simulations (for each FIRE-2 simulation, there is a DM-only one), the DM distribution within a few kpc exhibits higher density, with baryonic feedback in some simulations 
tending to produce core-like profiles~\cite{McKeown:2021sob}. 
These simulations are well suited for comparison with the Milky Way 
and are used to generate synthetic surveys resembling Gaia DR2
in data structure, magnitude limits, and observational
errors~\cite{2020ApJS..246....6S}. Interestingly, determination of the Milky Way mass profile using the Gaia DR2 galactic rotation curve and modelling of the baryonic components in the GC shows enhanced DM density in the inner 20 kpc from DM contraction induced by the presence of baryons~\cite{Cautun:2019eaf}. 
The FIRE-2 simulations exhibit a steep profile in the inner kpc. However, the inner 3 kpc densities in the FIRE-2 halos show a large spread, with sometimes baryonic feedback producing a core-like profile.
The central DM densities and velocity dispersions
are larger by a factor of $\sim$0.5 to 3, and $\sim$2.5–to 4, respectively, compared to  DM-only simulations, which has important consequence both for $s$ and $p$-wave annihilation models. The minimum spatial resolution for the Fire-2 simulation is about 400 pc, or $\sim$ 3$^{\circ}$. In the present work, we have extrapolated the J-factor profile to the gravitational centre of the MW using a linear approximation. Fig.~\ref{fig:fire2Jfact} shows the Fire-2 DM density profile, as well as the J-factors. The left panel shows the DM density, and the right panel shows the J-factors for both the $s$-wave and $p$-wave scenarios. The solid lines represent the mean of the density and J-factors, while the blue shaded regions represent the spread between the highest and smallest values obtained in the simulations.

\subsection{Effective field theory}
\label{sec:EFT}
In this paper, we assume that the WIMP is a Dirac fermion $\chi$ with mass $m_\chi$ that is a singlet under the SM gauge group. Operators for possible DM-SM interactions have previously been presented up to mass dimension seven~\cite{Brod_2018}, with a comprehensive exploration of current constraints on the full set of interactions with quarks and gluons presented in Ref.~\cite{GAMBIT:2021rlp}. For the purposes of this study, we are instead primarily interested in exploring the potential reach of CTAO for a smaller, representative set of interactions. To this end, we focus on the following set of operators representing scalar, pseudoscalar, vector and axial-vector interactions 
\begin{eqnarray}
\label{eqn:EFT_opp1}
\mathcal{O}_S &=& \frac{m_q}{M_\star^3}(\bar\chi\chi )(\bar q q)\\
\label{eqn:EFT_opp2}
\mathcal{O}_P &=& \frac{m_q}{M_\star^3}(\bar\chi\gamma^5\chi) (\bar q\gamma^5 q),\\
\label{eqn:EFT_opp3}
\mathcal{O}_V &=& \frac{1}{M_\star^2}(\bar\chi\gamma^\mu\chi) (\bar q\gamma_\mu q),\\
\label{eqn:EFT_opp4}
\mathcal{O}_A &=& \frac{1}{M_\star^2}(\bar\chi\gamma^\mu\gamma^5\chi) (\bar q\gamma_\mu \gamma^5 q),
\end{eqnarray}
where $M_\star$ is the energy scale up to which the EFT is valid (roughly equal to the mediator mass in the UV completion of the theory, and assumed to be independent of quark flavour),  $m_q$ is the quark mass corresponding to the SM quark field $q$, and $\gamma_\mu,\gamma_5$ are the Dirac gamma matrices. The conventional factor of $m_q$ for the operators $\mathcal{O}_S$ and $\mathcal{O}_P$ is added because they have the same flavour structure as the SM quark mass terms. Such a structure is thus consistent with the principle of minimal flavour violation, present in beyond-SM physics models that survive low energy flavour constraints. 

\subsubsection{Indirect detection with gamma rays}
\label{sec:ind_detec}
In the DM EFT detailed above, the WIMP $\chi$ can annihilate with its antiparticle $\bar{\chi}$ to produce a flux of particles from regions of our universe rich in DM. For the CTAO, we will be interested in measurements of high energy gamma rays originating in the centre of the Milky Way. Assuming that $\chi$ and $\bar{\chi}$ are in a spherical DM halo of density $\rho_\chi$ with velocities $\overrightarrow{v_1}$ and $\overrightarrow{v_2}$, the flux of gamma rays on Earth is given by:
\begin{equation}
\begin{aligned}
\label{eq:dmflux}
\frac{d^2\Phi}{d E d\Omega}(E,\Omega)=
\frac{\langle \sigma v \rangle_n}{4\pi \epsilon\,2m_{\chi}^2}\sum_f  {\rm BR}_f \frac{d N_f}{d E}(E)\,\frac{dJ_n}{d\Omega}
\end{aligned}
\end{equation}
with 
\begin{equation}
%\begin{aligned}
\label{eq:jfactor}
\frac{dJ_n}{d\Omega} = \int_{\rm los}  d s\, \rho_{\chi}^{2}(r[s,\theta]) 
\times \int d^{3}v_{\chi} \,g_{r}(v_{\chi})\int d^{3}v_{\bar{\chi}}\, g_{r}(v_{\bar{\chi}}) \frac{|\overrightarrow{v_{\chi}} -\overrightarrow{v_{\bar{\chi}}}|^{n}}{c^n}\, ,
%\end{aligned}
\end{equation}
where $n$ = 0, 2 for $s$ and $p$-wave annihilations, respectively. 
$\langle \sigma v \rangle_n$ is the velocity-weighted annihilation cross section for the s or p-wave conntribution, and $dN_f/dE$ is the differential yield of gamma rays per annihilation in the channel $f$ with its branching ratio $BR_f$.
$\epsilon$ = 2 (4) for Majorana (Dirac) particles. This implicitly assumes that the $\chi$ and $\bar{\chi}$ number densities are equal. $\theta$ is the angle between the GC and the line of sight (los).
This formulation assumes that the $\chi$ velocity distribution $f(\overrightarrow{r},\overrightarrow{v})$ can be written as $f(\overrightarrow{r},\overrightarrow{v}) = \rho{(\overrightarrow{r})}g(\overrightarrow{v})$ normalized such that $\int d^3v f(\overrightarrow{r},\overrightarrow{v}) = \rho(\overrightarrow{r})$.
The radial distance $r$ from the GC is related to the distance along the los $s$ by $r = (s^2+r_\odot^2 - 2\,s\, r_\odot\,\rm cos\,\theta)^{1/2}$, where $r_\odot$ is the distance between the GC and the Sun taken to be r$_\odot$ = 8.1 kpc~\cite{GRAVITYdistanceGC}. The value of the DM density at the solar location is assumed to be $\rho(r_\odot)$ = 0.38 GeV cm$^{−3}$.

As we shall see shortly, $\langle \sigma v\rangle$ for  EFT operators will be composed of velocity-dependent and independent terms. 
This can be handled by rewriting Eq.~(\ref{eq:dmflux}) depending on 
 whether the annihilation cross section is independent of the
relative velocity $v$ of the self-annihilating DM particles (s-wave, \textit{i.e}, n = 0), or velocity-dependent (\textit{e.g.}, for p-wave n = 2, for d-wave n = 4):
\begin{equation}
\begin{aligned}
\label{eq:dmflux2}
\frac{d^2\Phi}{d E d\Omega}\Big|_n(E,\Omega)=
\frac{1}{16\pi\,m_{\chi}^2}\sum_{f}  {\rm BR}_{f} \frac{d N_{f}}{d E}\Big(A_{f}\frac{dJ_n}{d\Omega} \Big)\, \Big( \frac{v}{c} \Big)^{2n} , 
\end{aligned}
\end{equation}
where the terms $A_f$ are extracted from the expression for the annihilation cross sections under consideration as given below.

The value of the velocity-averaged annihilation cross section can be calculated for each of the operators in Eqns.~\ref{eqn:EFT_opp1} to~\ref{eqn:EFT_opp4}. Operators with a factor of $m_q$ are suppressed, particularly when annihilation to top quarks is not kinematically accessible. The annihilation rate is also expected to be reduced for the operators $\mathcal{O}_S$ and $\mathcal{O}_A$, for different reasons. For $\mathcal{O}_S$, the annihilation rate is $p$-wave suppressed and is proportional to the square of the DM velocity $v^2\approx 10^{-6}$,
\begin{equation}
\langle\sigma v\rangle_{\mathcal{O}_S}  =  \sum_q \Theta(m_\chi - m_q)\frac{m_q^2}{M_\star^6}\frac{3 m_\chi^2}{8 \pi} \left(1-\frac{m_q^2}{m_\chi^2}\right)^{3/2} \hspace{-5pt}v^2
\label{sigvOS},
\end{equation}
where $\Theta$ is the Heaviside step function, necessary to ensure that the final states of DM annihilation are kinematically accessible.
For $\mathcal{O}_A$ there is a helicity-suppressed $s$-wave term proportional to $m_q^2$ and an additional $p$-wave suppressed term,
\begin{equation}
\langle\sigma v\rangle_{\mathcal{O}_A} = \sum_q\Theta(m_\chi - m_q)\frac{1}{M_\star^4}\frac{m_\chi^2}{4 \pi}\sqrt{1-\frac{m_q^2}{m_\chi^2}}
%\nonumber\\
%&{}&
\times\left(6\frac{m_q^2}{m_\chi^2} +\frac{8-22 {m_q^2}/{m_\chi^2}+17 {m_q^4}/{m_\chi^4}}{4 \left(1-{m_q^2}/{m_\chi^2}\right)}v^2\right)%s,\nonumber\\
\label{sigvOA}.
\end{equation}
The annihilation rates for the $\mathcal{O}_P$ and $\mathcal{O}_V$ operators are given by 
\begin{eqnarray}
\langle\sigma v\rangle_{\mathcal{O}_P} &=& \sum_q\Theta(m_\chi - m_q)\frac{m_q^2}{M_\star^6} 
\frac{3 m_\chi^2}{16\pi} \sqrt{1-\frac{m_q^2}{m_\chi^2}} %\nonumber\\
%&{}&
\times\left(8+\frac{2-m_q^2/m_\chi^2}{1-m_q^2/m_\chi^2}v^2\right),\label{sigvOP}\\
\langle\sigma v\rangle_{\mathcal{O}_V} &=& \sum_q\Theta(m_\chi - m_q)\frac{1}{M_\star^4}\frac{m_\chi^2}{2\pi} \sqrt{1-\frac{m_q^2}{m_\chi^2}}%\nonumber\\
%s&{}&
\times\left( 6+3\frac{m_q^2}{m_\chi^2} +\frac{8-4 {m_q^2}/{m_\chi^2}+5 {m_q^4}/{m_\chi^4}}{8  \left(1-{m_q^2}/{m_\chi^2}\right)}v^2\right) ,\nonumber\\
&{}& \label{sigvOV},
\end{eqnarray}
where we note in particular that $\mathcal{O}_V$ is the only operator to have a fully unsuppressed $s$-wave term.

\subsubsection{Direct detection with underground experiments}
\label{sec:dd_eft}
Direct detection experiments aim to observe WIMPs via their interactions with nuclei or electrons in Earth-bound detectors, though here we only consider the case of nuclear scattering. The observed yield of events therefore depends on the WIMP-nucleon scattering rate, which in turn depends on which of the DM EFT operators is realised in Nature. For example, the operators $\mathcal{O}_A$ and $\mathcal{O}_P$ give rise to scattering rates that are suppressed due to the target nuclear spin $\vec{s}_N$ and/or the scattering momentum exchange $\vec{q}$. 
Direct detection limits are traditionally presented in terms of zero-momentum WIMP-nucleon cross-sections, derived from effective theories where the WIMP interacts with nucleons via either a scalar operator (SI) or an axial-vector operator (SD). To facilitate a meaningful comparison between these low energy limits and those computed for WIMP-quark interactions, it is necessary to relate the couplings of WIMP-quark operators to those of WIMP-nucleon operators. We perform this translation using a leading-order prescription as described in Ref.~\cite{DeSimone:2016fbz}.

The four WIMP-nucleon operators arising from the WIMP-quark operators considered in our analysis are:
\begin{align}
    c^{NS} O^{NS} &= c^{NS} \, \bar{\chi} \chi \, \bar{N} N, \label{eq:ONS} \\
    c^{NP} O^{NP} &= c^{NP} \, \bar{\chi} i\gamma^5 \chi \, \bar{N} i\gamma^5 N, \label{eq:ONP} \\
    c^{NV} O^{NV} &= c^{NV} \, \bar{\chi} \gamma^\mu \chi \, \bar{N} \gamma_\mu N, \label{eq:ONV} \\
    c^{NA} O^{NA} &= c^{NA} \, \bar{\chi} \gamma^\mu \gamma^5 \chi \, \bar{N} \gamma_\mu \gamma^5 N. \label{eq:ONA}
\end{align}
The coefficients \( c^{Ni} \) (with \( i = S, P, V, A \)) of these operators can be expressed in terms of the coefficients of the WIMP-quark effective operators. The expressions are as follows:
\begin{align}
    c^{NS} &= \frac{m_N}{M_{*}^3} \left( \sum_{q = u, d, s} f^{(N)}_{q} + \frac{2}{9} f^{(N)}_{G} \right), \label{eq:cNS} \\
    c^{NP} &= \frac{m_N}{M_{*}^3} \sum_{q = u, d, s} \left(1 - \frac{6 \bar{m}}{m_q} \right) \Delta^{(N)}_{q}, \label{eq:cNP} \\
    c^{NV} &= \frac{3}{M_{*}^2}, \label{eq:cNV} \\
    c^{NA} &= \frac{1}{M_{*}^2} \sum_{q} \Delta^{(N)}_{q}. 
    %\label{eq:cNA}
    \label{eqn:EFT_nucleoncouplings}
\end{align}
Here, \( M_{*} \) is the suppression scale of the higher-dimensional operators, \( m_N \) is the nucleon mass, \( m_q \) are the quark masses, and \( \bar{m} \) is defined as:

\begin{equation}
    \bar{m} = \left( \frac{1}{m_u} + \frac{1}{m_d} + \frac{1}{m_s} \right)^{-1}. 
    %\label{eq:mb}
    \label{eqn:barm}
\end{equation}
The quantities \( f^{(N)}_{q} \), \( f^{(N)}_{G} \), and \( \Delta^{(N)}_{q} \) are the nucleon form factors, which are experimentally measured parameters representing the scalar and axial-vector couplings between quarks and nucleons. The values of these form factors are taken from the default settings in the DarkSUSY package~\cite{bringmann2018darksusy}. We also display all relevant values in Table~\ref{tab:EFT_form_factors}. We take $N=$ neutron as often done as standard in the direct detection community due to enhanced sensitivity in Xenon from an odd number of neutrons in Xe$^{129/131}$.

\begin{table}[htbp]
\centering
\begin{tabular}{llll}
\toprule
\textbf{Symbol} & \textbf{Value} & \textbf{Units} & \textbf{Description} \\
\midrule
\( m_N \) & 0.939 & GeV & Neutron mass \\
$m_f$ & 246 & GeV & Higgs VEV \\
\( f_{u}^n \) & 0.019 & -- & \\
\( f_{d}^n \) & 0.041 & -- &  \\
\( f_{s}^n \) & 0.14 & -- &  \\
\( \Delta_u^n \) & \(-0.40\) & -- &  \\
\( \Delta_d^n \) & 0.77 & -- &  Quark-neutron form factors\\
\( \Delta_s^n \) & \(-0.12\) & -- &  \\
\( \tilde{f}^n_u \) & \(-0.44\) & -- &  \\
\( \tilde{f}^n_d \) & 0.84 & -- &  \\
\( \tilde{f}^n_s \) & \(-0.03\) & -- & \\
\( m_u \) & 0.0022 & GeV & Mass of \( u \) quark \\
\( m_d \) & 0.0047 & GeV & Mass of \( d \) quark \\
\( m_s \) & 0.096 & GeV & Mass of \( s \) quark \\
\( f_G \) & \( 1 - f_{u}^n - f_{d}^n - f_{s}^n \) & -- & Gluon scalar form factor \\
\bottomrule
\end{tabular}
\caption{Constants used for translating the direct detection constraints for the DM-nucleon interactions arising from the EFT and simplified models, as detailed in Secs.~\ref{sec:EFT} and \ref{sec:SM}.}
\label{tab:EFT_form_factors}
\end{table}

To compare these with experimental results, we need to compute the zero-momentum WIMP-nucleon cross-sections as used by direct detection experiments. A spin-independent (SI) cross section can be predicted from the scalar (\( O^{NS} \)) and vector (\( O^{NV} \)) operators, while a spin-dependent (SD) cross section can be predicted from the axial-vector operator (\( O^{NA} \)). The cross-sections are given by:
\begin{align}
    \sigma^{SI} &= \frac{\mu^{2}_{\chi N}}{\pi} \left( c^{Ni} \right)^2, \quad \text{for } i = S, V, \label{eq:sigmaSI} \\
    \sigma^{SD} &= \frac{3 \mu^{2}_{\chi N}}{\pi} \left( c^{NA} \right)^2, \label{eq:sigmaSD}
\end{align}
where \( \mu_{\chi N} = \frac{m_\chi m_N}{m_\chi + m_N} \) is the WIMP-nucleon reduced mass, and \( m_\chi \) is the WIMP mass.

\subsubsection{EFT validity for direct detection}
\label{sec:valifityDD}

{ The operators of an EFT must ultimately be the low energy approximations of some new theory that can only be properly observed at an appropriately high energy scale. Above the requisite energy scale, one would observe new particles that explicitly mediate the interactions between DM and SM fields, whilst in the EFT, the interactions appear as contact interactions. For an EFT to provide a valid description of any given experiment, the energy scale of the underlying model must be much higher than the energies involved in the interactions under study. As a rough rule of thumb, this requires that the mass scale $M_*$ up to which the EFT is assumed to be valid is well above the typical momentum exchange in any experiment which one is attempting to model with the EFT (where the mass and momentum are both expressed in natural units). In principle, the operators of the EFT may arise from different processes in the higher energy theory, with one or more different mediators contributing in each case. In our approach, we assume a common new-physics scale $M_*$ for all of the effective operators, even though they might originate from different mechanisms at the ultraviolet (UV) level and correspond to distinct scales. Essentially, $M_*$ should be regarded as the smallest among these scales—that is, the energy at which new particles or interactions first become relevant when going up in energy}. These new degrees of freedom may not impact all processes, so certain effective operators could remain accurate even at energies higher than $M_*$. However, since this cannot be ascertained from a low-energy perspective, we conservatively restrict the validity of the EFT to energies below $M_*$.

For the application of direct detection constraints, the primary requirement for $M_*$ is that it exceeds the hadronic scale, ensuring that the effective operators can be expressed in terms of free quarks and gluons. This condition is satisfied when $M_* \gtrsim 2\,\mathrm{GeV}$, which holds true for this study. Nonetheless, to evaluate direct detection constraints, one needs to determine the relic abundance of DM particles. This calculation relies on the cross-sections for processes like $\chi\chi \rightarrow q\bar{q}$ or $\chi\chi \rightarrow gg$. For the EFT framework to be applicable in this context, we require $M_* > 2 m_\chi$. Therefore, parameter points where $M_*$ is smaller than this threshold are considered invalid.

\subsection{Simplified models}
\label{sec:SM}
The EFT approach to WIMP model building can be extended through the use of \textit{simplified models} that add an explicit mediator between SM and DM fields, which extends the range of validity of the theory. Previous examinations of such models can be found in Refs.~\cite{Abdallah:2015ter,Arina:2018zcq,DeSimone:2016fbz,Albert:2017onk,Boveia:2016mrp,Kahlhoefer:2017dnp,Arcadi:2017kky, Morgante:2018tiq,DEramo:2016gos, Carpenter:2016thc, Abercrombie:2015wmb}, and comprehensive statistical fits of candidate models can be found in Refs.~\cite{Bagnaschi_2019,Chang:2022jgo,Chang:2023cki}. 

We assume once more that the WIMP is a gauge-singlet Dirac fermion. Then, to aid comparison with previous estimates of the CTAO reach for DM simplified models, we investigate the same four models that were studied in Ref.~\cite{Balazs:2017hxh}. These are chosen due to being the simplest realisations of the four effective operators described above, featuring a scalar, pseudoscalar, vector or axial-vector mediator

\begin{eqnarray}
\mathcal{L}_S &=& -g_\chi \phi \bar \chi \chi - \sum_q g_q \phi \frac{m_q}{m_{vev}} \bar q q, \\
\mathcal{L}_P &=& -i g_\chi \phi \bar \chi \gamma_5 \chi - \sum_q i g_q \phi \frac{m_q}{m_{vev}} \bar q \gamma_5 q, \label{eqn:simp_P}\\
\mathcal{L}_V &=& -g_\chi  Z'_\mu \bar \chi \gamma^\mu \chi - \sum_q g_q Z'_\mu \bar q\gamma^\mu q, \label{eqn:simp_V} \\
\mathcal{L}_A &=& -g_\chi Z'_\mu \bar \chi \gamma^\mu \gamma_5 \chi - \sum_q g_q Z'_\mu \bar q \gamma^\mu \gamma_5 q,
\end{eqnarray}
where %$m_q/m_f$
$m_q/m_{vev}$
arises from a Yukawa coupling written in terms of the Higgs vacuum expectation value %$m_f\approx 246$~GeV 
$m_{vev}\approx$ 246~GeV. Note that these models necessarily have more free parameters than the EFT case, and a decision needs to be made on how to present future constraints from a hypothetical CTAO analysis. Our first simplification is to assume that $g_q$ is the same for all flavours of quark. Then we follow the LHC DM Working Group in assuming the benchmark couplings strengths of $g_q=0.25,g_\chi=1$ for the V and A models, and $g_q=1,g_\chi=1$ for the S and P models~\cite{Boveia:2016mrp}. This allows limits to be set in the plane of the mediator and WIMP masses.

Indirect detection and direct detection processes follow the same logic as in Sec.~\ref{sec:EFT}, except we must replace the calculations of the annihilation and scattering rates.

\subsubsection{Indirect detection with gamma rays}

The total annihilation cross sections for each simplified model operator are given by 
%\begin{widetext}
{
\allowdisplaybreaks
\begin{eqnarray}
\langle\sigma v\rangle_{S}^\mathrm{total} = \langle\sigma v\rangle_{S}^{gg} +  \Theta(m_\chi - M_{\rm med}) \langle\sigma v\rangle_{S}^{SS} %\nonumber\\
&{}&+ \sum_q \Theta(m_\chi - m_q) \langle\sigma v\rangle_{S}^{q \bar q},\label{sigvtotS}\\
\langle\sigma v\rangle_{P}^\mathrm{total} = \langle\sigma v\rangle_{P}^{gg} +  \Theta(m_\chi - M_{\rm med}) \langle\sigma v\rangle_{P}^{PP} %\nonumber\\
&{}&+ \sum_q \Theta(m_\chi - m_q) \langle\sigma v\rangle_{P}^{q \bar q},\label{sigvtotP}\\
\langle\sigma v\rangle_{V}^\mathrm{total} = \Theta(m_\chi - M_{\rm med}) \langle\sigma v\rangle_{V}^{VV} %\nonumber\\
&{}&+ \sum_q \Theta(m_\chi - m_q) \langle\sigma v\rangle_{V}^{q \bar q},\label{sigvtotV}\\
\langle\sigma v\rangle_{A}^\mathrm{total} = \Theta(m_\chi - M_{\rm med}) \langle\sigma v\rangle_{A}^{AA} %\nonumber\\
&{}&+ \sum_q \Theta(m_\chi - m_q) \langle\sigma v\rangle_{A}^{q \bar q}.\label{sigvtotA}
\end{eqnarray}
%\end{widetext}
where all relevant contributions from annihilations to quarks, mediators and gluons are summed. Taking these in reverse order, in all models WIMPs can annihilate to quarks through the mediator, with the annihilation rates given by

%\begin{widetext}
{
\small
\allowdisplaybreaks
\begin{eqnarray}
\langle\sigma v\rangle_{S}^{q \bar q} &=& \frac{3 g_q^2 g_\chi^2}{8 \pi} \frac{m_q^2}{m_f^2} \frac{m_\chi^2 \left(1-\frac{m_q^2}{m_\chi^2}\right)^{3/2}}{\left(M_{\rm med}^2-4 m_\chi^2\right)^2+\Gamma_{S,\rm tot} ^2 M_{\rm med}^2}v^2,\label{sigvqqS}\\
 \langle\sigma v\rangle_{P}^{q \bar q} &=& \frac{3 g_q^2 g_\chi^2}{2 \pi} \frac{m_q^2}{m_f^2}  \frac{m_\chi^2 \sqrt{1-\frac{m_q^2}{m_\chi^2}}}{\left(M_{\rm med}^2-4 m_\chi^2\right)^2+\Gamma_{P, \rm tot} ^2 M_{\rm med}^2},\nonumber\\
 &{}& +\mathcal{O}(v^2 )\label{sigvqqP}\\
\langle\sigma v\rangle_{V}^{q \bar q} &=&  \frac{3 g_q^2 g_\chi^2}{2\pi} \frac{ \left(m_q^2+2 m_\chi^2\right)\sqrt{1-\frac{m_q^2}{m_\chi^2}}}{ \left(M_{\rm med}^2-4 m_\chi^2\right)^2+\Gamma_{V,\rm tot} ^2 M_{\rm med}^2}+\mathcal{O}(v^2 ),\label{sigvqqV}\\
\langle\sigma v\rangle_{A}^{q \bar q} &=& \frac{g_q^2 g_\chi^2}{2\pi}   \nonumber\\
&{}&\times\frac{3m_q^2 \left(1-\frac{4 m_\chi^2}{M_{\rm med}^2}\right)^2 \sqrt{1-\frac{m_q^2}{m_\chi^2}}+m_\chi^2 v^2+\mathcal{O}(m_q^2 v^2 )}{\left(M_{\rm med}^2-4 m_\chi^2\right)^2+\Gamma_{A,\rm tot} ^2 M_{\rm med}^2}\nonumber\\
&{}&+\mathcal{O}(v^4 ) \label{sigvqqA},
\end{eqnarray}
}
where $M_{\rm med}$ is the mass of the mediator and $\Gamma_{N,\rm tot}$ is the total decay width of the mediator for the model $N$. Ignoring the possibility of leptonic final states for simplicity, this is given by the sum of the individual decay widths for each allowed final state

%\begin{widetext}
{
\allowdisplaybreaks
\begin{eqnarray}
\Gamma_{S,\rm tot} &=& \Gamma_S^{gg} + \sum_q \Theta(M_{\rm med} - 2m_q) \Gamma_S^{q \bar q}\nonumber\\
&{}&+\Theta(M_{\rm med} - 2m_\chi)\Gamma_S^{\chi \bar \chi},\\
\Gamma_{P,\rm tot} &=& \Gamma_P^{gg} + \sum_q \Theta(M_{\rm med} - 2m_q) \Gamma_P^{q \bar q}\nonumber\\
&{}&+  \Theta(M_{\rm med} - 2m_\chi)\Gamma_P^{\chi \bar \chi},\\
\Gamma_{ V,\rm tot} &=&\sum_q \Theta(M_{\rm med} - 2m_q) \Gamma_V^{q \bar q}+\Theta(M_{\rm med} - 2m_\chi)\Gamma_V^{\chi \bar \chi},\nonumber\\
\\
\Gamma_{A,\rm tot} &=&\sum_q \Theta(M_{\rm med} - 2m_q) \Gamma_A^{q \bar q} +\Theta(M_{\rm med} - 2m_\chi)\Gamma_A^{\chi \bar \chi}\nonumber.
\end{eqnarray}
}
%\end{widetext}
%
The factors $\Gamma_A^{\chi \bar \chi}$ and $\Gamma_A^{q \bar q}$ are calculated using the expression for the decay width into fermions $f\in\{\chi,q\}$ for each model,
\begin{eqnarray}
\Gamma_{S}^{f\bar f} &=& c_f^{SP}  \frac{g_f^2 M_{\rm med}}{8 \pi} \left(1-\frac{4m_f^2}{M_{\rm med}^2}\right)^{3/2}, \label{widthS}\\
\Gamma_{P}^{f\bar f} &=& c_f^{SP}  \frac{g_f^2 M_{\rm med}}{8 \pi} \sqrt{1-\frac{4m_f^2}{M_{\rm med}^2}},  \\
\Gamma_{V}^{f\bar f} &=& c_f^{VA}  \frac{g_f^2 M_{\rm med}}{12 \pi} \sqrt{1-\frac{4m_f^2}{M_{\rm med}^2}} \left(1+\frac{2m_f^2}{M_{\rm med}^2}\right),  \\
\Gamma_{A}^{f\bar f} &=&  c_f^{VA}  \frac{g_f^2 M_{\rm med}}{12 \pi} \left(1-\frac{4m_f^2}{M_{\rm med}^2}\right)^{3/2} ,\label{widthA}
%Checked against 1603.04156
\end{eqnarray}
where $c_\chi^{SP}=c_\chi^{VA} = 1$, $c_q^{SP} = 3 m_q^2 / m_f^2$, and $c_q^{VA} = 3$. In the $S$ and $P$ models, one must also include the relevant decay width to gluons
\begin{eqnarray}
\Gamma_{S}^{gg} &=& \frac{ \alpha_S^2 g_q^2 M_{\rm med}^3 \left| f_{\rm scalar}\left(\frac{4 m_t^2}{M_{\rm med}^2}\right)\right| ^2}{32 \pi ^3 m_f^2},\\
\Gamma_{P}^{gg} &=& \frac{ \alpha_S^2 g_q^2 M_{\rm med}^3 \left| f_{\rm pseudoscalar}\left(\frac{4 m_t^2}{M_{\rm med}^2}\right)\right| ^2}{32 \pi ^3 m_f^2}, \label{widthPgg}
\end{eqnarray}
expressed in each case in terms of either the $f_{\rm scalar}$ or $f_{\rm pseudoscalar}$ function
{
\allowdisplaybreaks
\begin{eqnarray}
f_{\rm scalar}(\tau) = \tau  \left(1+(1-\tau ) \tan^{-1}\left(\frac{1}{\sqrt{\tau -1}}\right)^2\right) ,\\
f_{\rm pseudoscalar}(\tau) = \tau \tan^{-1}\left(\frac{1}{\sqrt{\tau -1}}\right)^2.
\end{eqnarray}
}
These more complicated functional forms arise from the fact that the decay to gluons occurs via a quark loop.

The annihilation rates to mediators are given by
\allowdisplaybreaks
\begin{eqnarray}
\langle\sigma v\rangle_{S}^{SS} &=& \frac{g_\chi^4 m_\chi v^2}{24 \pi  }\nonumber\\
&{}&\times\frac{\left(9 m_\chi^6-17 M_{\rm med}^2 m_\chi^4+10 M_{\rm med}^4 m_\chi^2-2 M_{\rm med}^6\right)}{\sqrt{m_\chi^2-M_{\rm med}^2} \left(2 m_\chi^2-M_{\rm med}^2\right)^4},\nonumber\\
&{}&\label{sigvSS}\\
\langle\sigma v\rangle_{P}^{PP} &=&\frac{g_\chi^4 m_\chi \left(m_\chi^2-M_{\rm med}^2\right)^{5/2}}{24 \pi  \left(2m_\chi^2-M_{\rm med}^2\right)^4}v^2 ,\label{sigvPP}\\
\langle\sigma v\rangle_{V}^{VV} &=&\frac{g_\chi^4 \left(m_\chi^2-M_{\rm med}^2\right)^{3/2}}{4 \pi  m_\chi \left(M_{\rm med}^2-2 m_\chi^2\right)^2}+\mathcal{O}(v^2 ),\label{sigvVV}\\
\langle\sigma v\rangle_{A}^{AA} &=&\frac{g_\chi^4 \left(m_\chi^2-M_{\rm med}^2\right)^{3/2}}{4 \pi  m_\chi \left(M_{\rm med}^2-2 m_\chi^2\right)^2}+\mathcal{O}(v^2 ) \label{sigvAA},
\end{eqnarray}
where the process can only occur if the WIMP mass exceeds the mediator mass, and the mediators will subsequently decay into SM particles.

Finally, for the $S$ and $P$ models, we must include the annihilation rate to gluons given by 

\allowdisplaybreaks
\begin{eqnarray}
\langle\sigma v\rangle_{S}^{gg} &=&\frac{\alpha^2}{8 \pi^3 m_f^2}%
\frac{2 g_q^2 g_\chi^2 m_\chi^4 v^2}{\left(M_{\rm med}^2-4 m_\chi^2\right)^2+\Gamma_{S,\rm tot} ^2 M_{\rm med}^2}\nonumber\\
&{}&\times\left|\sum_q f_{\rm scalar}\left(\frac{m_q^2}{m_\chi^2}\right)\right| ^2  ,\label{sigvggS}\\
\langle\sigma v\rangle_{P}^{gg} &=&\frac{\alpha^2}{2 \pi^3 m_f^2}%
\frac{g_q^2 g_\chi^2}{\left(M_{\rm med}^2-4 m_\chi^2\right)^2+\Gamma_{P,\rm tot} ^2 M_{\rm med}^2}\nonumber\\
&{}&\times\left|\sum_q m_q^2 f_{\rm pseudoscalar}\left(\frac{m_q^2}{m_\chi^2}\right)\right| ^2 \label{sigvggP}. 
\end{eqnarray}

\subsubsection{Direct detection}
\label{sec:dd_simp}
The pseudoscalar interaction leads to a spin-dependent, momentum-suppressed scattering process. Assuming that the mediator mass \( M_{\text{med}} \) is appropriately large, the Lagrangian for the pseudoscalar mediator results in the following expression for the scattering cross section~\cite{PhysRevD.83.115009}:

\begin{equation}
\sigma^{\text{SD}} = \frac{9}{4} \frac{f_N^2}{m_\chi^2 m_N^2} \frac{g_\chi^2 \mu_{\chi N}^2}{M_{\text{med}}^4} q^4,
\label{eq:sd_cross_section}
\end{equation}
where \( q \) is the momentum transfer, \( N = n, p \) denotes nucleons (neutrons or protons), \( m_\chi \) is the DM mass, \( m_N \) is the nucleon mass, \( \mu_{\chi N} = \frac{m_\chi m_N}{m_\chi + m_N} \) is the reduced mass of the DM–nucleon system, and \( g_\chi \) is the DM coupling to the mediator. The factor \( f_N \) encapsulates the nucleon matrix elements and is given by~\cite{dienes2014overcoming}:
\begin{equation}
f_N = m_n \left[ \sum_{q = u, d, s} g_q \frac{m_f}{m_q} \tilde{f}_q - \bar{m} \left( \sum_{q = u, d, s} \tilde{f}_q \frac{m_q}{m_f} \right) \sum_{q = u, d, s, c, b, t} g_q \frac{m_f}{m_q} \right],
\label{eq:f_N}
\end{equation}
with $\bar{m}$ is given by Eqn.~(\ref{eqn:barm}) and $mf\simeq 246$ GeV is the Higgs vacuum expectation value, \( m_n \) is the nucleon (again neutron) mass, \( g_q \) are the mediator couplings to quarks, \( m_q \) are the quark masses and \( \tilde{f}_q \) are pseudoscalar form factors for quarks \( q = u, d, s \). The $\tilde{f}_q$ are adapted from Refs.~\cite{Balazs:2017hxh} and have  values also displayed in Tab.~\ref{tab:EFT_form_factors}. The expression in Eq.~\eqref{eq:sd_cross_section} can be directly compared to the limits reported by direct detection experiments. However, the momentum suppression (due to the \( q^4 \) factor) causes the spin-dependent scattering cross section to be orders of magnitude below current experimental sensitivity. For couplings of order unity and mediator masses above a few GeV, even the next generation of direct detection experiments will not impose restrictive limits on the parameter space of this simplified model.

In contrast, for the vector mediator case, the DM–nucleon scattering is spin-independent and not velocity suppressed. The vector current interaction is proportional to the number of valence quarks in the nucleon. The resulting scattering cross section is given by~\cite{PhysRevD.83.115009,ALiamFitzpatrick_2013}:
\begin{equation}
\sigma^{\text{SI}} = \frac{9 \mu_{\chi N}^2 g_\chi^2 g_q^2}{\pi M_{\text{med}}^4}.
\label{eq:si_cross_section}
\end{equation}
Since the analysis assumes \( g_\chi = 1 \) and \( g_q = 0.25 \), the scattering cross-section is relatively large, necessitating heavy mediators to evade the direct detection limits.

\subsubsection{Unitarity limit}
The simplified model framework presumes the existence of an unspecified ultraviolet (UV) completion that ensures the internal consistency of the model, including mechanisms for mass generation and the preservation of gauge invariance. However, in the case of the axial-vector mediator model, certain processes can violate gauge invariance and perturbative unitarity within specific regions of the parameter space. In these regions, any UV completion would significantly alter the phenomenology of the model~\cite{Kahlhoefer:2015bea}. To prevent violations of perturbative unitarity, the following condition must be satisfied:
\begin{equation}
\begin{aligned}
m_{\chi} &\lesssim \sqrt{\frac{\pi}{2}} \, \frac{M_{\text{med}}}{g_{\chi}}\;,
\end{aligned}
\label{eq:unitarity_conditions}
\end{equation}
where $m_\chi$ is the DM mass, $m_q$ is the quark mass, $M_{\text{med}}$ is the mediator mass, and $g_\chi$, $g_q$ are the couplings to DM and quarks, respectively.

\section{Experimental constraints on dark matter models}
\label{sec:signals}

In this section, we provide details of our implementation of a hypothetical future CTAO indirect DM search, alongside our treatment of existing limits from direct search experiments and the Large Hadron Collider.

\subsection{Sensitivity forecast with the CTAO}
\label{sec:analysis}
%\EM{@Igor: please mention the hard energy threshold}
The Cherenkov Telescope Array Observatory (CTAO) is a forthcoming array of Imaging Atmospheric Cherenkov Telescopes (IACT) consisting of a two-site observatory with an array in each hemisphere. It will be sensitive to gamma-rays of energies in the range of 20 GeV - 300 TeV, and will improve the sensitivity over current generation IACTs by a factor up to 10, while also increasing the angular resolution by a factor of up to 2~\cite{CTAConsortium:2017dvg}. 

Our target of observations is the Galactic Center (GC) region, and therefore we will make use of the Southern-hemisphere array of CTAO, as its location in Chile enables to observe this region under the best conditions. The southern one (CTAO-South) is planned to be at Paranal in Chile (-24$^{\circ}$S 37$'$, -70$^{\circ}$W 24$'$) and expected to be composed of 
14 medium-sized telescopes and 37 small-sized telescopes, known as the \textit{Alpha configuration}. The current publicly-available Instrument Response Functions (IRFs) \texttt{prod5-v0.1} are taken from the \texttt{prod5-v0.1} dataset~\cite{CTAprod5} with the \texttt{Performance-prod5-v0.1-South-20deg} scheme optimized for 50 hour observation of a point-like source at a zenith angle of 20${^\circ}$.

To perform our analysis, we consider a circular region of 5$^{\circ}$ in radius centered on the GC as our observational target for CTAO, with a 500 hour homogeneous time exposure over the whole region. This exposure for the GC is consistent with CTAO's expected survey of the GC region~\cite{CTAConsortium:2017dvg}. This observation region is further divided into Regions of Interest (ROIs), created as 47 concentric annuli of 0.1$^{\circ}$ width around the GC~\cite{HESS:2022ygk}. The residual background, effective area, angular and energy resolutions considered here are obtained from the above-mentioned IRFs of the CTAO-S site. For this analysis, we consider an energy range from 50 GeV up to 140 TeV.

Cosmic rays interacting with interstellar
gas, magnetic, and radiation fields produce diffuse
gamma-ray emission. In the GeV energy range, this Galactic Diffuse Emission (GDE) dominates the overall flux, and a significant contribution in the TeV energy range is expected. Current models for the spatial and spectral TeV-scale GDE assume different realistic CR source densities and interstellar radiation field distributions.
Despite their differences, the models considered agree with an extensive collection of locally measured CR data and are
expected to be representative of the gamma-ray uncertainties
related to the transport of CRs in the Galaxy. 
Here we consider a realistic model of the expected GDE 
including the $\pi_0$ and inverse Compton scattering (ICS) components from the most recent release of the CR propagation framework GALPROP (version 57)~\cite{Porter:2021tlr}. We make use of the SA100
(R12) model as a baseline model for the GDE predictions with GALPROP extracted from Ref.~\cite{Montanari:2023sln}. 
A detailed study of the effects of different GDE models on the sensitivities to EFT/SM DM models is out of the scope of
the present study. We will show how much the sensitivity can be improved when the GDE is subtracted from the measured background for the CTAO observations. The expected energy-differential flux for our chosen GDE model is shown in Fig.~\ref{fig:EFT_flux1TeV}. 

Additionally, different VHE gamma-ray sources are located in the GC. These sources include HESS J1745-290~\cite{Aharonian:2009zk}, the supernova remnant HESS J1745-303~\cite{Aharonian:2008gw}, a diffuse emission correlated with clouds of the Central Molecular Zone~\cite{Abramowski:2016mir}, as well as standard VHE sources in the Galactic plane~\cite{HESS:2018pbp}. In order to avoid complex modelling of each of these sources in a coherent GC background model, we place a set of masks to cover bright and extended emissions~\cite{HESS:2022ygk} and exclude a band of Galactic latitudes between $\pm0.3^{\circ}$, covering the Galactic plane.

In what follows, we make use of the Fire-2 DM density simulation J-factors computed in Ref.~\cite{McKeown:2021sob} for $s$ and $p$-wave contributions to the WIMP annihilation plotted in Fig.~\ref{fig:fire2Jfact}, calculating them in the whole 5$^\circ$ ROI around the GC~\footnote{We note that an alternative approach to J-factors for velocity-dependent annihilation cross sections (based on the Eddington inversion formula) is presented in Ref.~\cite{Boucher:2021mii}, but we do not pursue this further in our study.}. Given the suppression of the $p$-wave component in many of our models, a proper treatment of the J-factors for velocity-dependent scattering is crucial to determine the correct CTAO reach. Fig.~\ref{fig:EFT_flux1TeV} shows %in solid lines
the expected VHE gamma-ray flux in the observation region for the 4 EFT operators as well as for the 4 simplified model mediators, for a DM mass of 1 TeV. We also show, for comparison, the expected CTAO residual background (dashed-dotted line) and the model for the GDE background (dashed line) used in  this work.
\begin{figure}
     \hspace{-0.5cm}
    \includegraphics[scale = 0.3]{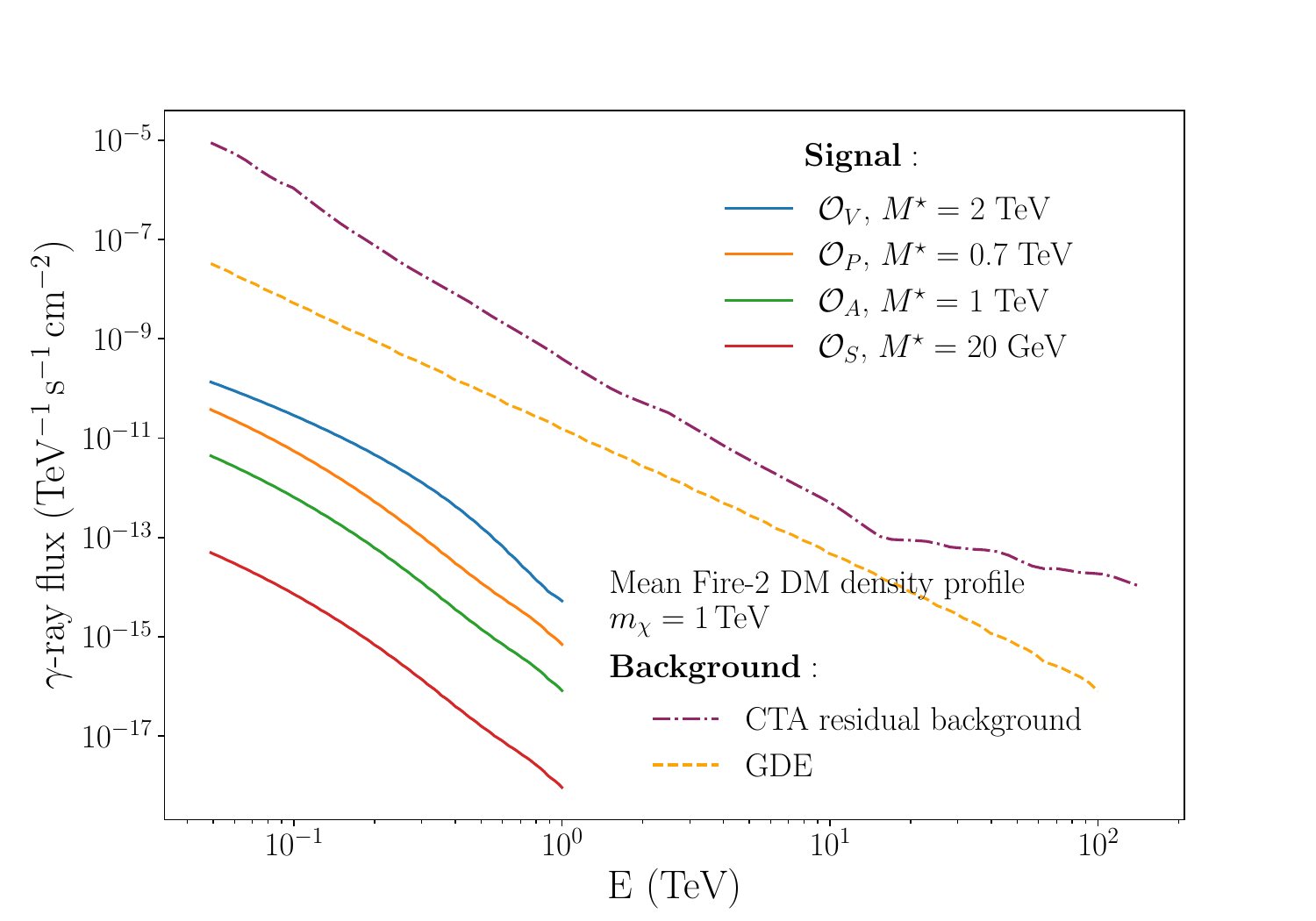}
    \hspace{-0.5cm}
    \includegraphics[scale = 0.3]{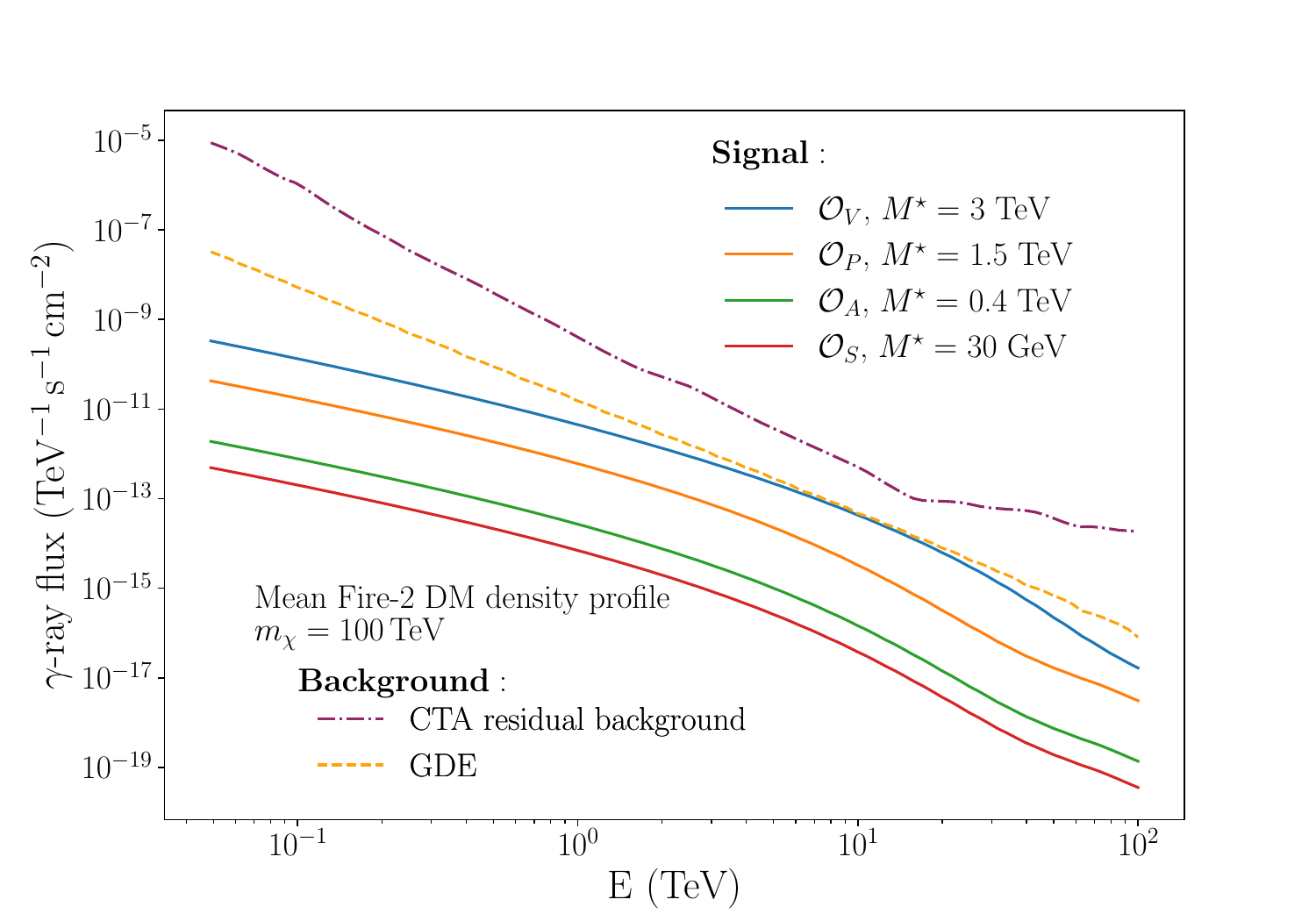}
      \hspace{-0.5cm}
    \includegraphics[scale = 0.3]{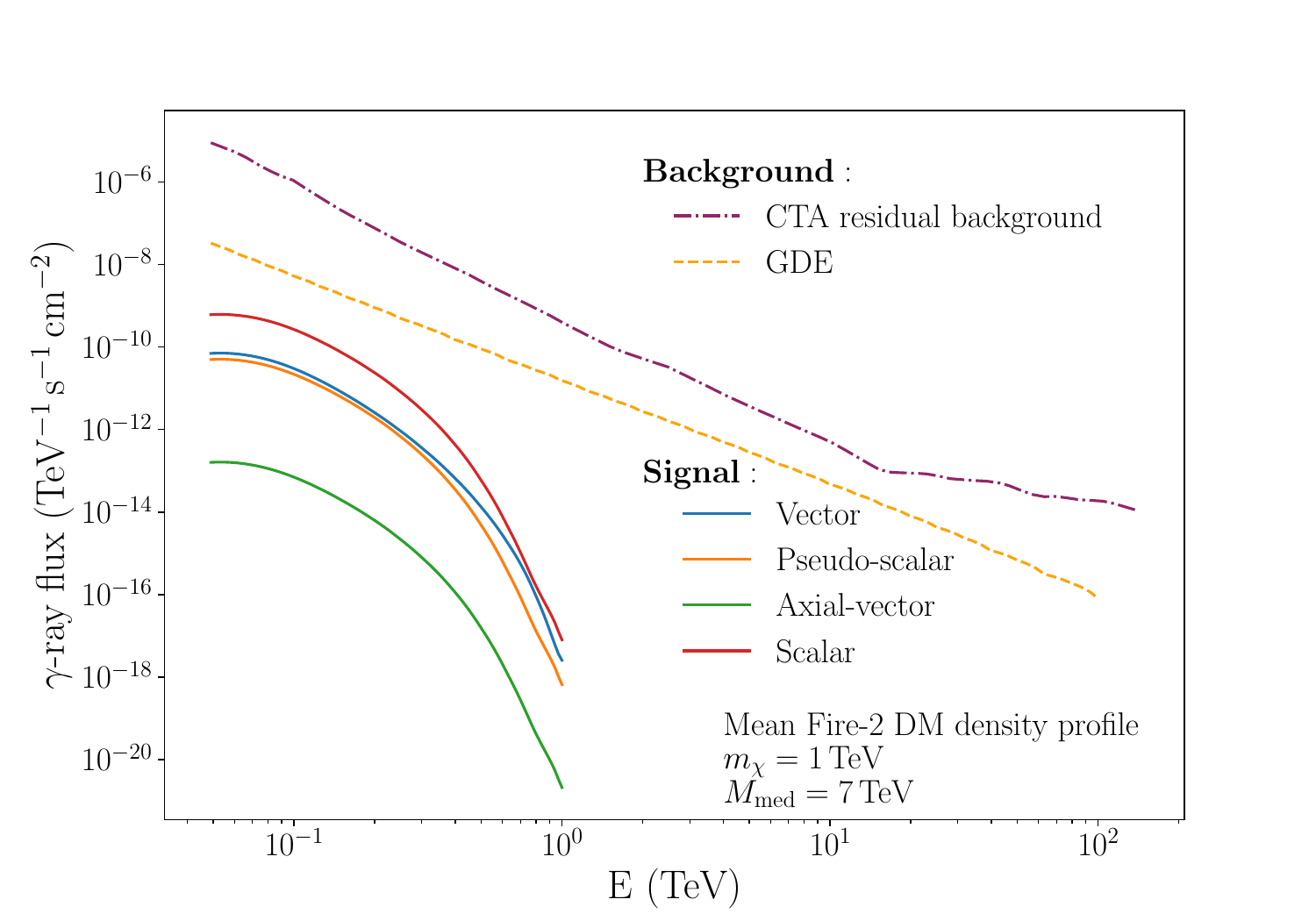}
    \hspace{-0.cm}
    \includegraphics[scale = 0.3]{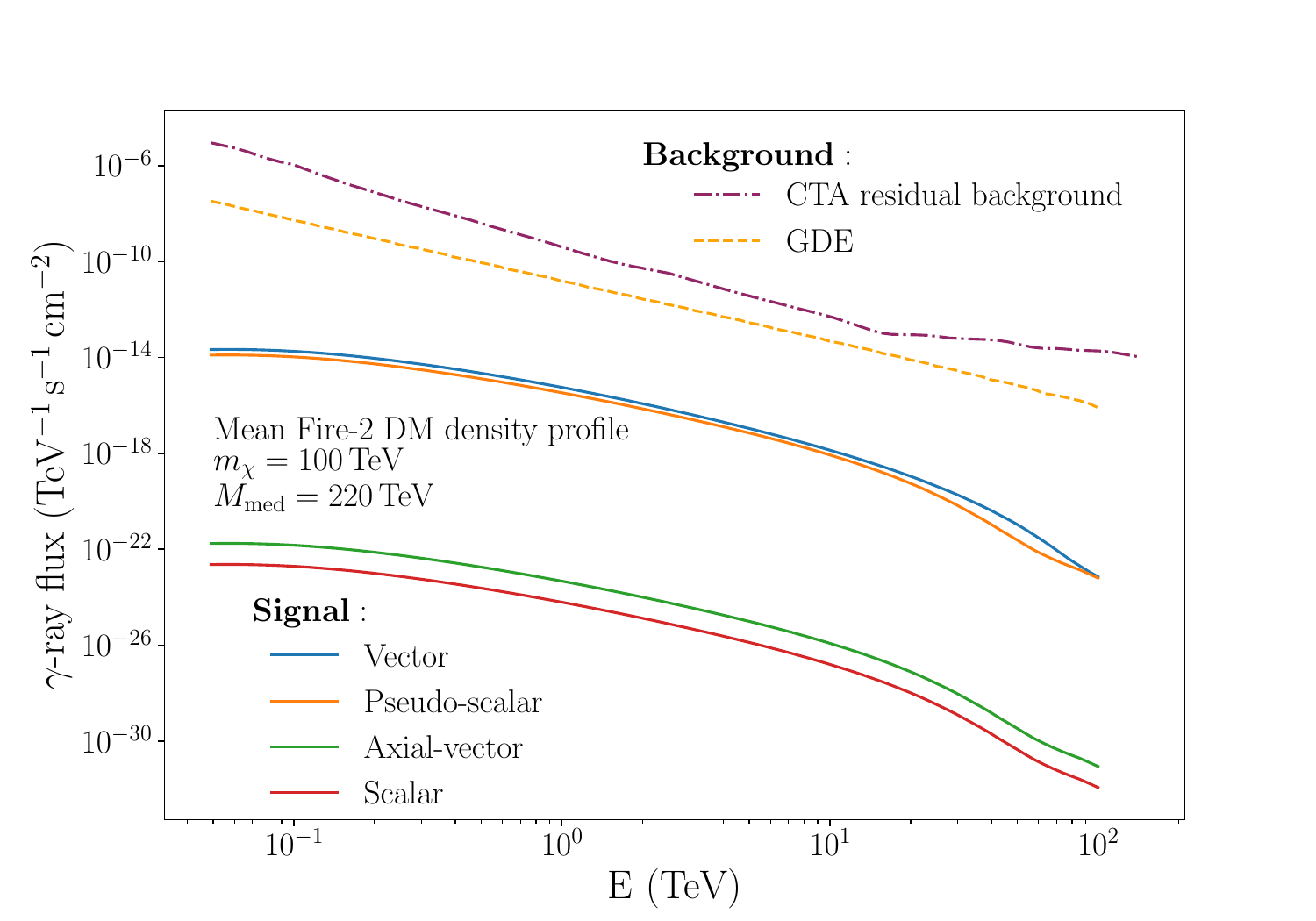}
    
    \caption{
    Energy-differential VHE gamma-ray flux expected in the EFT (top panels) and Simplified Model (bottom panels) approaches, respectively, for 1 TeV DM mass (left) and 100 TeV DM mass (right). The J-factor is taken as the mean value obtained from the 12 profiles discussed in Ref.~\cite{McKeown:2021sob}.
    The fluxes are integrated over the inner 5$^\circ$ of the GC excluding the masked regions.
    The expected fluxes are plotted for the four EFT operators  representing scalar ($\mathcal{O}_S$), pseudoscalar ($\mathcal{O}_P$), vector ($\mathcal{O}_V$) and axial-vector ($\mathcal{O}_A$) interactions, with different values of $M_*$. 
    For the simplified models, the expected fluxes are plotted for four effective operators featuring a scalar, pseudoscalar, vector or axial-vector mediators and two mediator masses, respectively. Also shown are the VHE gamma-ray backgrounds integrated in the same region as for the expected signals, including the
    CTAO residual background flux  (dashed-dotted violet line)  as well as our adopted GDE flux (orange dashed line).
     }
    \label{fig:EFT_flux1TeV}
\end{figure}

The count number from the expected flux in a given ROI $i$ and energy bin $j$ with central value $E_j$ is given by:
\begin{equation}
    N^S_{ij} = T_{{\rm obs}, i} \int_{E_j - \Delta E_j /2}^{E_j + \Delta E_j /2} dE \int_{-\infty}^{\infty} dE'\, \frac{d\Phi^S_{ij}}{dE'} (\Delta \Omega_i, E')\, A_{\rm eff}^{\gamma}(E')\, G(E - E') \, ,
\label{eq:count}
\end{equation}
where $A_{\rm eff}^{\gamma}$ and $G$ are the energy-dependent effective area and energy resolution, taken from the CTAO IRFs, and  $T_{\rm obs}$ is the total observation time, taken as 500 hours.
Similarly, the expected count number from the residual background can be obtained by substituting the gamma-ray flux $d\Phi_{ij}^{S}/dE'$ with the background flux $d\Phi_{ij}^{B}/dE'$ and performing the same calculations using Eq.~(\ref{eq:count}).

We ultimately wish to impose constraints to $M_{*}$, and to do so we need to perform a statistical analysis. The expected number of signal counts $N^{S}_{ij}$ shown in Eq.~\ref{eq:count} is binned, calculated in the \textit{j}th energy and \textit{i}th spatial bins. The data is divided in 92 logarithmically-spaced energy bins between 50 GeV and 140 TeV $\Delta E_{j}$, and spatial bins of solid angle $\Delta\Omega_{i}$ composed by the different annuli ROIs. Similarly, we also divide the expected background count $N^{B}_{ij}$ in energy and spatial bins.

We perform the statistical analysis and computation of the expected sensitivity using a log-likelihood ratio test statistic (TS)~\cite{Cowan:2011an}. Using both the spectral and spatial characteristics of the expected signal, we employ an ON-OFF method, where the ROIs are taken as the ON regions, and the OFF regions are taken as control regions, positioned sufficiently far from the GC, used to estimate the expected background at the ROIs, and having the same angular size as the ON regions. The flux considered for the OFF regions is taken from Monte Carlo simulations of the residual background and extracted from the CTAO IRFs~\cite{CTAprod5}. The likelihood is also binned in two dimensions, one energetic and one spatial, and can be written as

\begin{eqnarray}
\mathcal{L}_{ij} (N_{ij}^S, N_{ij}^B, \bar{N}_{ij}^S, \bar{N}_{ij}^B\, |\, N_{ij}^{\rm ON}, N_{ij}^{\rm OFF})  
%& = &
=\textrm{Pois}[(N^S_{ij} + N^B_{ij}), N^{\rm ON}_{ij}] \nonumber \\ 
\times \textrm{Pois}[(\bar{N}^S_{ij} + \bar{N}^B_{ij}), N^{\rm OFF}_{ij}] 
\label{eq:likelihood}
%\end{equation}
\end{eqnarray}
 where $\textrm{Pois}[\lambda,n] = e^{-\lambda}\lambda^{n}/n!$. $N^{ON}_{ij}$ and $N^{OFF}_{ij}$ are the numbers of events observed in the ON and OFF regions, binned in \textit{i} spatial and \textit{j} energetic bins. $N^{S}_{ij}$ and $\bar{N}^{S}_{ij}$ represent the expected signal in the ON and OFF regions, respectively. Similarly, $N^{B}_{ij}$ and $\bar{N}^{B}_{ij}$ quantify the background in the ON and OFF regions. Since the residual background used in this analysis comes from Monte Carlo simulations, we can say that the signal in the OFF regions is zero, and therefore $\bar{N}^{S}_{ij} = 0$.

To calculate the TS we need a full likelihood, and to obtain it 
from Eq.~(\ref{eq:likelihood}) we must simply compute the product of each binned likelihood $\mathcal{L}_{ij}$, such as $\mathcal{L} =  \prod_{\rm ij} \mathcal{L_{\rm ij}}$. For a DM mass $m_{\chi}$, we can set $M^{*}$ as a free parameter in the TS, and we define the TS as 
\begin{equation}
    \text{TS}(m_{\chi}) = - 2 \ln \frac{\mathcal{L}(M^{*}, m_{\chi})}{\mathcal{L}(\widehat{M^{*}}, m_{\chi})},
    \label{eq:TS}
\end{equation}
\noindent where $\widehat{M^{*}}$ denotes the value of $M^*$ that maximises the likelihood for a DM mass of $m_{\chi}$. In the limit of large statistics TS follows a $\chi^{2}$ distribution with a  single degree of freedom. Under this  condition we can set one-sided 95\% upper limits on $M^*$ by solving TS for the coupling above the best fit with TS = 2.71. To do so, the dataset is generated using the Asimov procedure~\cite{Cowan:2011an}. We compute the sensitivity treating the mean expected background as data. The sensitivity is then computed as the mean expected lower limit, at 95\% confidence level (C.L.). For the SM case, one should substitute $M^*$ by $M_{\rm med}$ by analogy in the TS computation. Note that in this case, $g_q$ and $g_\chi$
are set to given values.

\subsection{Constraints from Xenon1T and LZ experiments}
\label{sec:dd}
We extract the 2018 Xenon1T~\cite{aprile2018dark} and 2024 LUX-ZEPLIN (LZ)~\cite{aalbers2024dark} direct detection constraints %via the re-casting 
on the zero momentum WIMP-nucleon (neutron in the case of SD) scattering cross-sections $\sigma_{SI}/\sigma_{SD}$ versus dark matter mass. %\cite{XENON:2018voc}. 
The expressions used to recast these limits into EFT and simplified model parameters
%to perform the re-casting 
depend on the EFT/simplified model framework and on the effective operator nature (S,P,V,A). This recasting procedure makes use of the publicly available \texttt{DDCalc} code~\cite{GAMBIT:2021rlp}.

Of primary concern in this study is the change in the direct detection constraints arising from the updated assumptions on local astrophysical parameters that are relevant for direct detection. In order to correctly account for the scaling of these limits in light of the Fire-2 simulations, we extract the minimum and maximum 3D velocity dispersions from the Fire-2 simulations and re-calculate the associated direct detection limit with \texttt{DDCalc}~\cite{GAMBIT:2021rlp} under the assumption that changes to the associated local DM density $\rho_\text{loc}$ and velocity dispersion $\sigma_{3D}$ from the Fire-2 simulations will scale the collaboration limits similarly to using the full extended Poisson likelihood function utilised by the XENON 
collaboration~\cite{XENON:2024xgd}. We then estimate the multiplicative difference between the official direct detection limit and that produced by \texttt{DDCalc} with the updated Fire-2 astrophysical parameters, and use this factor to adjust the rescaled collaboration limits.

Direct detection experiments are designed to observe the scattering of dark matter (DM) particles from the Galactic halo off nuclei in an ultra-pure target by measuring the recoil energy \(E_{\mathrm{R}}\) of the struck nucleus.  The differential rate of nuclear recoils per unit target mass and per unit recoil energy is given by
\begin{equation}
\frac{\mathrm{d}R}{\mathrm{d}E_{\mathrm{R}}}
= \frac{\rho_\text{loc}}{m_{T}\,m_{\chi}}
\int_{v_{\min}(E_{\mathrm{R}})}
v\,f(v)\,\frac{\mathrm{d}\sigma}{\mathrm{d}E_{\mathrm{R}}}\,\mathrm{d}^{3}v \,,
\end{equation}
where \(\rho_\text{loc}\) denotes the local DM density, \(m_{T}\) is the mass of the target nucleus, \(m_{\chi}\) is the DM particle mass, and \(f(v)\) is the local DM speed distribution.  The lower limit of the velocity integral,
\begin{equation}
v_{\min}(E_{\mathrm{R}})
= \sqrt{\frac{m_{T}\,E_{\mathrm{R}}}{2\,\mu^{2}}}\,,
\end{equation}
corresponds to the minimum DM speed required to impart a recoil energy \(E_{\mathrm{R}}\), with \(\mu = m_{T}\,m_{\chi}/(m_{T} + m_{\chi})\) being the reduced mass of the DM–nucleus system. The distribution of velocities $\mathbf{v}$ of dark matter particles in the Earth's frame is given by

$$
f(\mathbf{v}, t)=\tilde{f}\left(\mathbf{v}_{\mathrm{obs}}(t)+\mathbf{v}\right)
$$
where the velocity $\mathbf{v}_{\mathrm{obs}}(t)$ is the velocity of the Earth
in the halo, and the velocity distribution in the galactic rest frame $\tilde{f}$ is assumed to be Maxwellian in nature:
\begin{align}
\label{eqn:halo}
\tilde{f}(\mathbf{v})\propto\left(\pi v_0^2\right)^{-3 / 2} e^{-\mathbf{v}^2 / v_0^2}
\end{align}
where $v_0$ is the most probable speed.

We consider two extreme (min/max) values for $\sigma_{3D}$ extracted from the Fire-2 simulations at a radius of $r$ = 8.3 kpc. These are $\sigma^\text{min}_{3D} = 278$ km/s and $\sigma^\text{max}_{3D} = 336$ km/s respectively. These correspond to a most probable speed~\cite{Beniwal:2022rde,Savage:2006qr} of $v_0^\text{min} = 227$ km/s and  $v_0^\text{max} = 274$ km/s in Eqn.~\ref{eqn:halo}. For the local DM density we adopt the Fire-II result of $\rho_\text{loc} = 0.38$ GeV / cm$^3$. Numerical units were handled by the Natpy package \cite{Howson:2021rph}. In order to correctly estimate the effect of these parameters via rescaling, we first take the astrophysical parameters ($\rho_\text{loc},\;v_0$, $\mathbf{v}_{\mathrm{obs}})$  adopted by the XENON and LZ collaborations (Refs.~\cite{XENON:2018voc} and \cite{aalbers2024dark} respectively), and then recalculate the limits solely changing $\rho_\text{loc}$ and $v_0$ to match the Fire-II results as prescribed above. We observe a maximum factor of $\sim 5\times$ increase in sensitivity for both SI and SD limits at $m_\chi\sim10$ GeV when using the Fire-2 parameters with $\sigma^\text{max}_{3D}$. This difference decreases to zero for $m_\chi\gtrsim50$ GeV.

\subsection{LHC constraints from ATLAS and CMS}
\label{sec:lhc_eft}
The effective operators introduced in Sec.~\ref{sec:EFT} enable the production of WIMP pairs in proton-proton collisions at the LHC. When one of the incoming partons emits a jet via initial state radiation (ISR), the process \( pp \rightarrow \chi \chi j \) can be observed as a monojet event associated with missing transverse energy \(\boldsymbol{E}_T\). In our analysis,  we use the Global and Modular Beyond-the-Standard-Model Inference Tool (\texttt{GAMBIT}~\cite{GAMBIT:2017yxo}), to incorporate the monojet studies from CMS~\cite{CMS:2017zts} and ATLAS~\cite{ATLAS:2021kxv}, which are based on data samples of \(36~\mathrm{fb}^{-1}\) and \(139~\mathrm{fb}^{-1}\) from Run II, respectively (see Ref.~\cite{GAMBIT:2021rlp} for further details). Although ATLAS and CMS have conducted additional searches involving other ISR types, such as mono-photon signatures, these are known to impose weaker constraints on DM EFTs compared to monojet searches~\cite{Brennan:2016xjh,Bauer:2017fsw}.

In such LHC monojet searches for DM, the validity of the EFT approach necessitates that the invariant mass of the DM pair produced in collisions satisfies \( m_{\chi\chi} < M_* \)~\cite{Berlin:2014cfa}. To obtain reliable constraints, it is important to include only events with energy transfers below this scale when calculating likelihoods within \texttt{GAMBIT}. However, a challenge arises because \( m_{\chi\chi} \) is not directly observable—unlike the missing transverse energy \( \boldsymbol{E}_T \) of the event—making it difficult to evaluate how varying $M_*$ impacts the predicted LHC spectra. One potential solution is to generate new LHC events for each parameter point and include only those events where \( m_{\chi\chi} \) is sufficiently small in the likelihood computation. Unfortunately, this method is computationally impractical. In this work, we adopt a simpler strategy also adopted by the \texttt{GAMBIT} 
collaboration~\cite{GAMBIT:2021rlp}: instead of comparing $M_*$ to the invariant mass \( m_{\chi\chi} \), it is related to the typical energy scale of the event, estimated by the amount of missing transverse energy produced. Specifically, we do not alter the missing energy spectrum for \( \boldsymbol{E}_T < M_* \) and apply the EFT validity requirement only for larger values of \( \boldsymbol{E}_T \). This approach is less conservative than that advocated in Refs.~\cite{racco2015robust,bruggisser2016last}, where the energy scale of the event is taken to be the partonic center-of-mass energy \( \sqrt{{s}} \). However, it has the significant advantage that it can be implemented after event generation, since the differential cross-section with respect to missing energy, \( \mathrm{d}\sigma / \mathrm{d}\boldsymbol{E}_T \), is the quantity directly compared to experimental data. We acknowledge that other methods for recasting LHC limits after event generation have also been proposed for example in Refs.~\cite{Busoni_2014,Berlin_2014}, both on which Refs.~\cite{racco2015robust,bruggisser2016last} are based. We note that this prescription could be regarded as optimistic, as it can overestimate the EFT-valid region whenever the typical momentum transfer $q$ exceeds $M_\ast$. In particular, the mapping $q \sim \boldsymbol{E}_T$ neglects the dependence on the unknown UV couplings that enter $M_\ast = M_{\rm med}/\sqrt{g_q g_\chi}$, and may therefore lead to artificially strong limits for some choices of $g_q g_\chi$ or for events with very large $\boldsymbol{E}_T$. A full event-by-event EFT-validity recast is beyond the scope of this work, but we expect it would weaken our LHC bounds by factors of order unity within  the phase space considered, without affecting the complementarity conclusions presented here.

In this study, we consider a realistic approach is to implement a smooth cut-off that results in a non-zero but rapidly decreasing missing energy spectrum above $M_*$.  For this, we make the replacement:
\begin{equation}
\frac{\mathrm{d}\sigma}{\mathrm{d}\boldsymbol{E}_T} \rightarrow \frac{\mathrm{d}\sigma}{\mathrm{d}\boldsymbol{E}_T} \left( \frac{\boldsymbol{E}_T}{M_*} \right)^{-a},
\end{equation}
for \( \boldsymbol{E}_T > M_* \). Here, \( a \) is a free parameter dependent on the specific UV completion. The limits \( a \rightarrow 0 \) and \( a \rightarrow \infty \) correspond to no truncation and an abrupt cut-off above \( M_* \), respectively. In the case where the EFT arises from the exchange of an \( s \)-channel mediator with mass near \( M_* \), one finds \( a \approx 2 \)~\cite{buchmuller2014beyond}. Instead of being guided by a specific UV completion, we keep \( a \) as a nuisance parameter in the interval \([0,4]\) and determine the value that best fits the data at each parameter point. This method generally leads to conservative LHC bounds, as much stronger exclusions could be obtained in specific UV completions if the heavy particles generating the effective DM interactions can be directly produced at the LHC.

In the simplified model case, constraints from the LHC can  arise either from WIMP production involving the mediator, or from the production and decay of the mediator to standard model species. The specific characteristics of the DM model determine which dataset provides the strongest limits. For models involving a pseudoscalar mediator, monojet searches yield the most stringent constraints. We take the monojet constraints from ATLAS \cite{aad2021search}  and CMS ~\cite{tumasyan2021search} experiment  for $\sqrt{s}=13$ TeV and a total integrated luminosity of 139 and $137\;\text{fb}^{-1}$ respectively  with $g_q=g_\chi=1$.  In contrast, for models with vector mediators, mono-X searches are no longer the most effective. Instead, searches for dijet resonances with large invariant masses provide greater sensitivity~\cite{abdallah2015simplified}. We use the ATLAS dijet constraints at an integrated luminosity of $139\;\text{fb}^{-1}$~\cite{ATLAS:2024kpy}. CMS do not have a dijet analysis of simplified DM models, and so we present the analogous constraints from the  monojet  study of \cite{tumasyan2021search}.

\section{Results and discussion}
\label{sec:results}
We now present and discuss the results of our analysis, covering the EFT and simplified model cases in turn.

\begin{figure}[ht!]
\hspace{-1.75cm}
    \includegraphics[scale = 0.75]{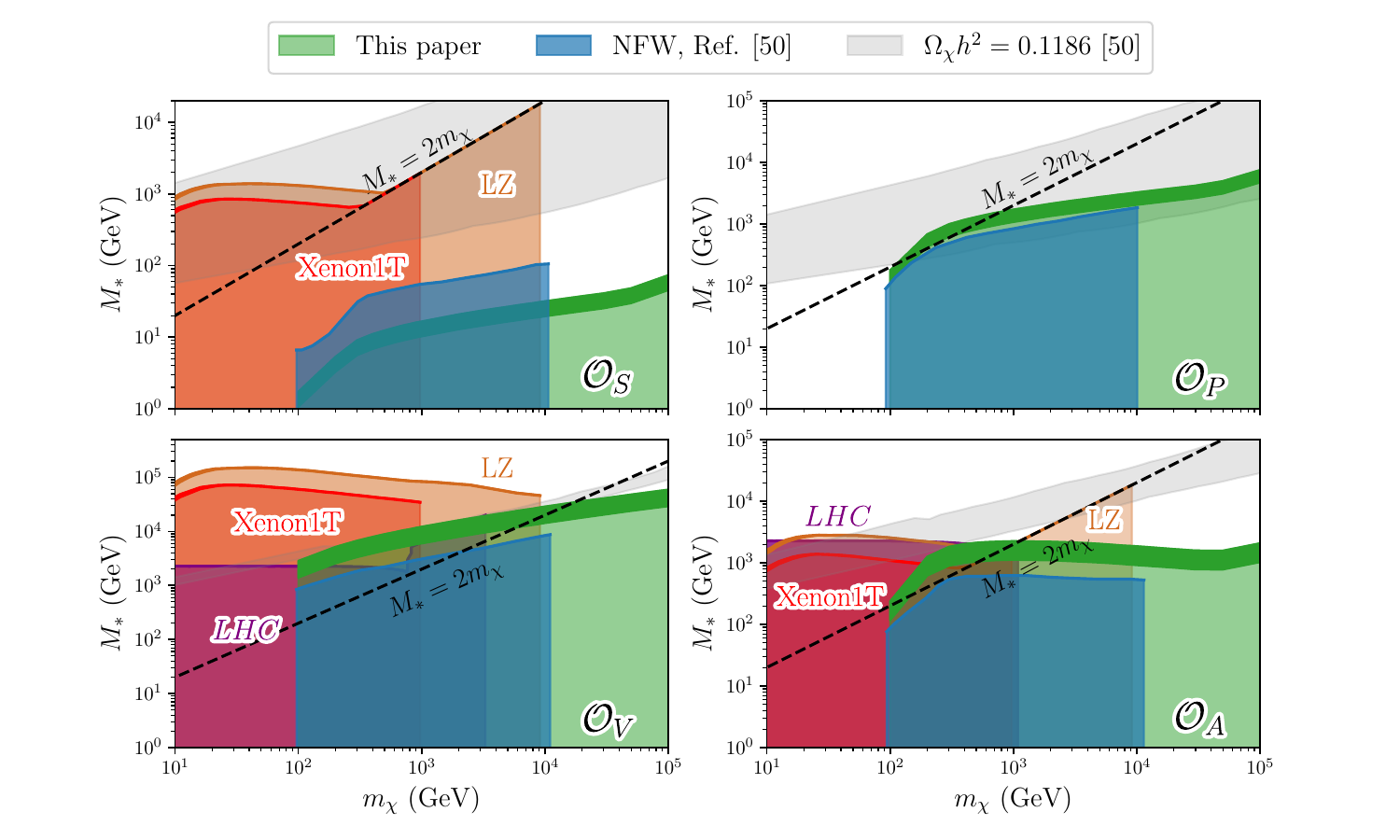}
    \caption{
    CTAO sensitivity for scalar (top left), pseudo-scalar (top right), vector (bottom left) and axial-vector (bottom right) EFT operators expressed in the (m$\chi$, M$_*$) plane. The sensitivity is expressed as the 95\% C. L. mean expected lower limit.
    The thickness of the green line represents the impact of the lowest and highest J-factors. 
    The red and orange-shaded areas show 90\% C.L. lower limits derived from Xenon1T \cite{XENON:2018voc} and LZ \cite{aalbers2024dark} measurements, respectively. The 95\% C. L. lower limit from a combination of the CMS~\cite{CMS:2017zts} and ATLAS~\cite{ATLAS:2021kxv} monojet/mono-X searches is shown by the violet line. 
    The thickness of the red and orange lower limits are obtained considering the minimum and maximum velocity dispersion values from the distribution given in Ref.~\cite{McKeown:2021sob}. The blue-shaded regions show previous sensitivities extracted from Ref.~\cite{Balazs:2017hxh} assuming a DM distribution following a NFW profile. Finally, we display the contour $M_* = 2m_\chi$ as the black dashed line, for which only the region above the dashed line is kinematically available. The gray shaded region represents the region of the parameter space that is able to retrieve the DM relic abundance $\Omega_{\chi}h^{2} = 0.1186$.
    }
    \label{fig:CTA_comp}
\end{figure}
\subsection{Effective field theory}
\label{sec:subsecEFT}
Fig.~\ref{fig:CTA_comp} shows the CTAO sensitivity for each EFT operator in the $m_\chi-M_*$ plane, where points above the black dashed line satisfy the EFT validity condition $M_* > 2m_\chi$. As the velocity dispersion changes, so does the J-factor and the expected signal given by Eq.~\ref{eq:dmflux}. Varying the velocity from its minimum to maximum value in the DM velocity dispersion distribution yields a spread in J-factors, and therefore in the sensitivity curve plotted as the dark green regions. Comparing our results  with those from Ref.~\cite{Balazs:2017hxh} shown as the blue-shaded region, we show that, except for the Scalar model, we improve the expected sensitivity for CTAO due to the use of the up-to-date J-factors in the GC region. Note that Ref.~\cite{Balazs:2017hxh}
makes use of a NFW DM distribution both for s- and p-wave J-factors. In the $\mathcal{O}_S$ case, this has the consequence of producing worse results when we consider the Fire-2 simulations, simply because the Scalar operator is only dependent on the p-wave J-factor. 
The Fire-2 p-wave J-factor is orders of magnitude smaller than the NFW ones used in Ref.~\cite{Balazs:2017hxh}. For a DM particle with mass 1 TeV, we reach a maximum $M_{*}$ sensitivity of 16.6 GeV, 1.7$\times$10$^{3}$ GeV, 1.2$\times$10$^{4}$ GeV and 2.2$\times$10$^{3}$ GeV for the
$\mathcal{O}_S$, $\mathcal{O}_P$, $\mathcal{O}_V$ and $\mathcal{O}_A$
operators,  respectively. It is worth noting that the results from Ref.~\cite{Balazs:2017hxh} used a now obsolete CTAO telescope array configuration with an optimistic overall number of telescopes yielding a quite optimistic flux sensitivity. The gray-shaded region extracted from Ref.~\cite{Balazs:2017hxh} represents the region of the parameter space for which the relic DM density is compatible 
%with that is able to retrieve the
measured CDM one $\Omega_{\chi}h^{2} = 0.1186$. 
%and was also extracted from Ref.~\cite{Balazs:2017hxh}.
Such region is %The gray-shaded region indicates the parameter region for which the DM density c
computed assuming a standard thermal freeze-out. %is compatible the measured CDM density.
%with a standard thermal freeze-out, 
However, it is important to note that departures from the standard cosmological history, such as modified expansion rates~\cite{Bringmann:2021tjr}, prolonged reheating~\cite{Drewes:2017fmn}, long-lived heavy scalar fields in the early universe~\cite{Fukuda:2024ddb} or additional non-thermal production channels, can substantially alter the relic abundance. Therefore, the full parameter space probed here remains relevant, even when the standard thermal relic density does not saturate the value obtained from cosmological measurements.

Strictly speaking, above the electroweak scale the $\mathcal{O}_\mathcal{S}$ and $\mathcal{O}_\mathcal{P}$ operators correspond to higher-dimensional interactions involving the Higgs doublet, which also induce annihilation processes such as $\chi\chi \rightarrow t \bar{t} h$. We have checked that, in the mass range considered in Fig.~\ref{fig:CTA_comp}, the additional Higgs emission modifies the annihilation cross-section by at most $\mathcal{O}(0.1)$ factors compared to $\chi\chi \rightarrow t \bar{t}$, and does not qualitatively change our CTAO reach. We acknowledge that in the pseudoscalar case, a fully consistent UV completion typically requires an extended Higgs sector (i.e a second Higgs doublet), implying that very large $m_\chi$ and $M_*$ should be interpreted with appropriate caution. Additionally, we follow the  treatment of Ref.\cite{GAMBIT:2021rlp} and neglect renormalization group-induced mixing and running effects, which are sub-leading in the parameter region where the EFT is valid.

Fig.~\ref{fig:CTA_comp} also shows the limits on $M_{*}$ arising from the Xenon1T and LZ experiments. Published limits expressed in terms of $\sigma^{SI}$ or $\sigma^{SD}$ as functions of $m_\chi$, are translated into functions of the effective suppression scale $M_*$ by inverting Eqs.~\eqref{eq:sigmaSI} and \eqref{eq:sigmaSD}. For operators $\mathcal{O}_S$ and $\mathcal{O}_P$, interactions the LHC mono-jet sensitivity is strongly suppressed by their $m_q$ couplings. Given this suppression, no limits satisfying the EFT validity condition survive and are thus not shown. For the scalar and vector cases, DD offers the most stringent limits up to a mass of $m_\chi\sim10^4$ GeV, with LZ achieving the most sensitivity in $M^*$ for all operators considered, reaching $\sim10^3, 10^5$ and $5\times10^3$ GeV for the $\mathcal{O}_S,\mathcal{O}_V$ and $\mathcal{O}_A$ cases respectively. For the $\mathcal{O}_V$ and $\mathcal{O}_S$ cases, we additionally observe a sudden linear increase in the DD limits for $m_\chi\gtrsim10^3$ GeV corresponding  to the EFT validity $M_* > 2m_\chi$ being saturated in these regions.} 

Finally, the combined ATLAS/CMS constraints are observed to only contribute non-trivially for the vector and axial-vector cases due to the effect of the more realistic EFT considerations detailed in Sec. \ref{sec:lhc_eft}. We observe that the LHC limits exceed the CTAO limits for the case of the axial-vector operator for masses  $m_\chi<500$ GeV and both DD and CTAO limits for $m_\chi>10^3$ GeV. Conversely,  LHC limits remain subdominant to DD and CTAO constraints for the case of the vector operator. 

\subsection{Simplified models}

\begin{figure}
    \centering
    \includegraphics[width=1\linewidth]{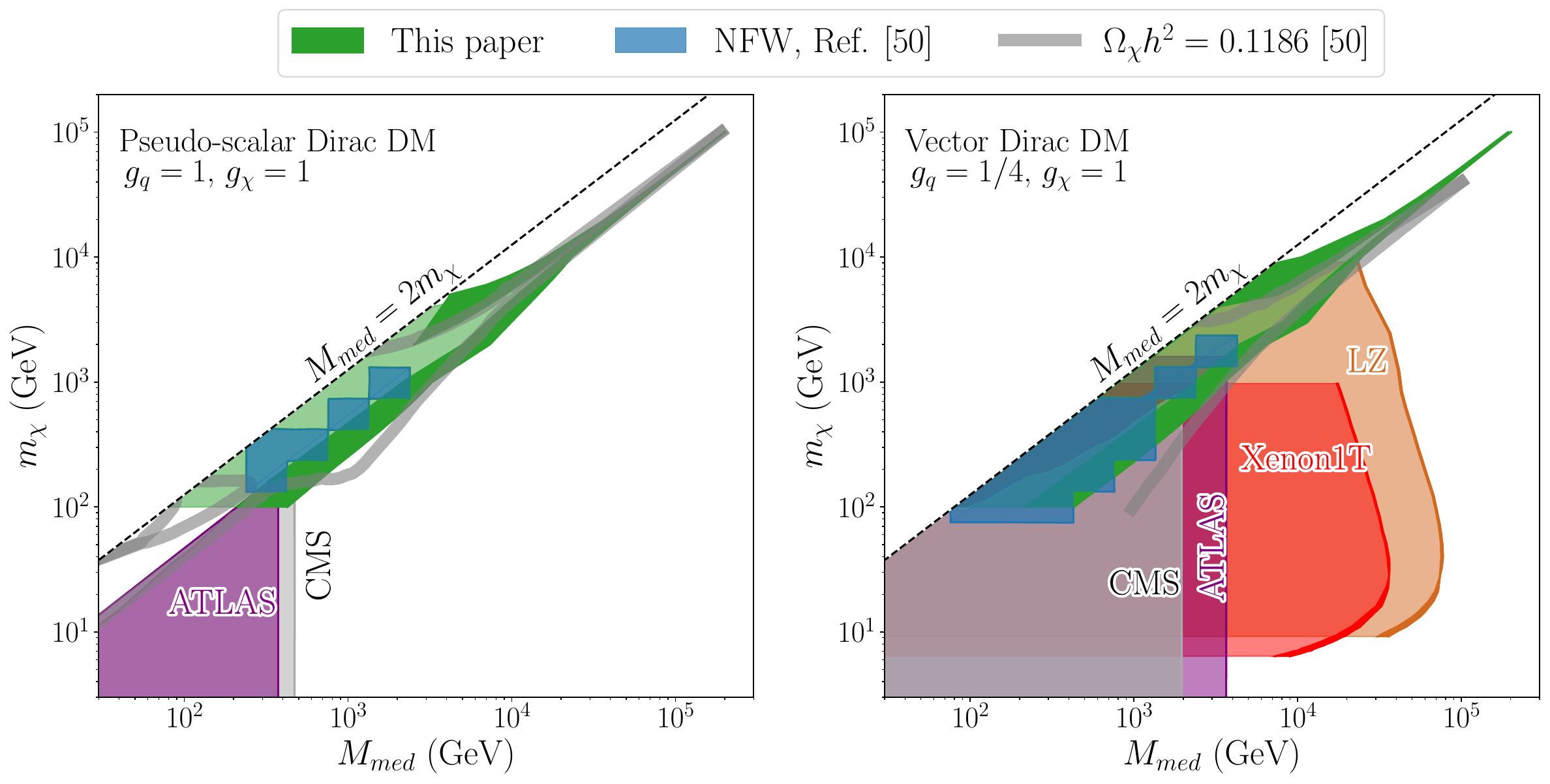}
    \caption{CTAO sensitivity (green shaded region) for pseudo-scalar (left) and  vector (right) SM operators expressed in the (M$_{\rm med}$, m$_\chi$) plane. The sensitivity is expressed as the 95\% C. L. mean expected lower limit. The dark shaded region shows the impact of the lowest and highest J-factors.
    The black dashed line corresponds to the contour of the kinematic constraint $M_{med} = 2m_\chi$, where only the region below the line is kinematically available.
    The red and orange shaded areas show 90\% C.L. lower limits from Xenon1T \cite{XENON:2018voc} and LZ \cite{aalbers2024dark}, respectively. The violet and grey regions represent the exclusion limits from the ATLAS and CMS monojet searches \cite{aad2021search,tumasyan2021search} for the pseudo-scalar case. As for the vector case, the purple region provides limits from ATLAS dijet 
     constraints~\cite{ATLAS:2024kpy} and the gray region shows limits from the CMS monojet search \cite{tumasyan2021search}. The thickness of red and orange lower limit curves represent the effect of considering the minimum and the maximum values of the dispersion velocity distribution given in Ref.~\cite{McKeown:2021sob}. The blue shaded squared regions show previous sensitivities extracted from Ref.~\cite{Balazs:2017hxh} assuming a DM distribution following a NFW profile. The gray shaded curves represents the region of the ($m_{\chi},\,M_{med}$) parameter space that is able to retrieve the DM relic abundance $\Omega_{\chi}h^{2} = 0.1186$.}
    \label{fig:SMpseudo}
\end{figure}

Fig.~\ref{fig:SMpseudo} shows the CTAO sensitivity for the SM in the (M$_{\rm med}$,m$_\chi)$ plane. Only the Pseudo-scalar and Vector DM particle models are displayed. This is due to the fact that for the Scalar and Axial-vector cases, CTAO would not have enough sensitivity to probe the relevant parameter space  for the expected signal: the CTAO respective sensitivity regions would be located above the unitarity limit curves, a region not allowed in these models. This is a consequence of the strong suppressions in the cross section in these two cases (see Eqs. \ref{sigvtotS} and \ref{sigvtotA}). The dark green regions show the difference in sensitivity due to the DM velocity dispersion uncertainty. Another noteworthy difference for the sensitivity shown here is the shape of the green region. We see an upper and lower bound, respectively, due to the energy dependence of the annihilation cross section, that presents a resonance. For small DM and mediator masses, CTAO is sensitive to a region around the resonance, and as masses increase, this region becomes narrower, to the point where only \textit{resonant masses} could produce a signal high enough to be detectable. For a DM particle of mass 1 TeV, we reach an upper sensitivity on the mediator mass $M^{*}$ of 2.5$\times$10$^{3}$ GeV and 2.9$\times$10$^{3}$ GeV for the pseudo-scalar and vector Dirac DM models, respectively. The gray bands represent the region of the parameter space that is able to retrieve the DM relic abundance $\Omega_{\chi}h^{2} = 0.1186$~\cite{Balazs:2017hxh}. 

Fig.~\ref{fig:SMpseudo} also shows constraints from Xenon1T and LZ  on DM–nucleon scattering mediated by pseudoscalar and vector mediators in the non-relativistic limit derived by the P and V Lagrangians in Eqn.~\ref{eqn:simp_P} and \ref{eqn:simp_V} respectively.  No points are excluded in the scalar Dirac DM case for $g_q = g_\chi = 1$ and none of the excluded parameters of the axial-vector model obey perturbative unitarity, and so these cases are not considered. XENON1T and LZ limits in the \( M_{\rm med}\)–\( m_\chi \) plane for the pseudoscalar and vector mediators are shown in red and orange respectively, where the region to the left of the lines are excluded. We conduct the analysis for the pseudoscalar mediator with \( g_\chi = 1 \) and \( g_q = 1 \), and for the vector case we take \( g_\chi = 1 \) and \( g_q = 0.25 \). For the Pseudoscalar case, the DD limits only achieve a maximal sensitivity of $M_{\rm med} \sim 0.1$ GeV due to the $q^4$ suppression in the cross-section \ref{eq:sd_cross_section}, and therefore are not visible on the left plot of Fig.~\ref{fig:SMpseudo}. In the case of the vector operator we observe a maximal sensitivity of $M_{\rm med}\sim 6\times 10^4$ GeV for a DM mass of $m_\chi\sim45$ GeV.

Finally, we consider LHC constraints on the simplified model scenarios. In the left panel of Fig.~\ref{fig:SMpseudo}, we display the monojet constraints from ATLAS \cite{aad2021search}  and CMS ~\cite{tumasyan2021search} experiment  for $\sqrt{s}=13$ TeV and a total integrated luminosity of 139 and $137\;\text{fb}^{-1}$ respectively  with $g_q=g_\chi=1$. While these searches impose strong limits at low DM masses, their sensitivity diminishes rapidly for $m_\chi\gtrsim500$ GeV. For the vector mediator case, we display the  ATLAS dijet constraints for an integrated luminosity $139\;\text{fb}^{-1}$~\cite{ATLAS:2024kpy} in the right panel of Fig.~\ref{fig:SMpseudo}. By applying stringent cuts on the invariant mass of dijet events, it is possible to suppress the substantial background from quantum chromodynamics (QCD) and effectively search for vector mediators. These searches are therefore particularly sensitive to the mediator-quark coupling $g_q$ and the mediator mass. As a result,  the limits are relatively independent of the DM mass for fixed $g_q=0.25,\;g_\chi=1$. CMS do not have a dijet analysis of simplified DM models, and so we present the analogous constraints from the  monojet  study of \cite{tumasyan2021search}.

\section{Summary and conclusions}
\label{sec:conclusions}

We performed in this work a realistic sensitivity study of annihilating WIMP models in the context of EFT and simplified model approaches,
for CTAO with 500 hours of observations in the inner 5$^\circ$ of the GC.
We make use of the most up-to-date IRFs of CTAO-South and
incorporate a model for the Galactic diffuse emission. Up-to-date DM density profiles, J-Factors and velocity dispersion distributions as relevant for s- and p-wave annihilations have also been considered from the Fire-2 hydrodynamical simulations. Direct detection constraints have been updated accordingly. The sensitivity reached in this study provided improved constraints for the $M_{*}$ and $M_{med}$ parameters in the EFT case and in the SM one, respectively, as a function of the WIMP mass when compared to previous studies. We were also able to cover a region of the parameter space at high $m_{\chi}$ values for which previous CTAO studies have indicated no sensitivity. These improvements can be seen in Figs.~\ref{fig:CTA_comp} and \ref{fig:SMpseudo}. The complementarity between indirect detection with gamma rays, direct detection and LHC searches has also been updated according to the latest limits for the latter two. 
Each experimental endeavour covers the parameter space differently, leading to a greater combined sensitivity in overlapping regions.

In some cases, as for the Pseudo-scalar EFT operator, the LHC and direct detection constraints are not present, and therefore we are left to rely only on the CTAO indirect detection sensitivity which allows the exploration of uncharted region of the parameter space. {For the pseudoscalar operator, no direct detection limits can be derived due to the EFT condition being violated as discussed in Secs.~\ref{sec:valifityDD} and \ref{sec:subsecEFT}. Furthermore, as discussed in Sec.~\ref{sec:subsecEFT},  LHC limits on the EFT constraints are observed to only contribute non-trivially for the vector and axial-vector cases due to the effect of the more realistic EFT considerations detailed in Sec.~\ref{sec:lhc_eft}. In the axial vector case, the LHC monojet/mono-X searches offer stronger constraints than the CTAO projection for intermediate WIMP masses, though CTAO will extend the reach to higher masses. In the vector case, the LHC limits are always subdominant to the projected limit from CTAO where the CTAO limit exists. In both of these case, regions of overlapping sensitivity offer exciting possibilities for \textit{combined} studies of the particle physics of DM should Nature have selected those particular couplings. 

In the simplified models cases, no points are excluded in the scalar Dirac DM case for $g_q = g_\chi = 1$ and none of the excluded parameters of the axial-vector model obey perturbative unitarity, and so these cases are not considered in our study. Additionally, DD limits only achieve a maximal sensitivity of $M_{\rm med} \sim 0.1$ due to  each $\gamma_5$ matrix in the Lagrangian leading to a momentum suppression in the WIMP-nucleon scattering cross-section (Eqn.\ref{eq:sd_cross_section}). For the Scalar and Vector operators however, we see that Xenon1T and LZ constraints dominate the parameter space, and the CTAO forecast only complements their results, showing advantage only when probing higher DM masses (see Fig.~\ref{fig:CTA_comp}). Similarly, for the Pseudo-scalar simplified model case, CTAO can probe a large portion of the parameter space that is currently unavailable at the LHC, going up to masses of 10$^{5}$ GeV. The LHC constraints, which are dominated by the monojet search channel, have only a small region of overlap with the CTAO projected limits. In the simplified model Vector case, however, a significant portion of the sensitivity region of CTAO is complementary to LHC and direct detection constraints. Only in the high-mass resonance region does it probe a new unexplored region of the parameter space (see Fig. \ref{fig:SMpseudo}). Again, this offers exciting options for complementary studies of the particle physics of DM.

With the advent of CTAO, observations of the GC in VHE gamma rays will play a crucial role in complementing the reach of direct detection and collider searches and probing yet uncharted regions of the parameter spaces for WIMP DM models. CTAO will therefore allow for novel synergies with direct detection and collider experiments in the pursuit of particle DM.

\section*{Acknowledgements}
This work was conducted in the context of the CTAO Consortium. It has made use of the CTAO instrument response functions provided by the CTAO Consortium and Observatory.
The authors would like to thank Felix Kahlhoefer for useful conversations regarding the EFT formalism and development of the code used to generate the LZ constraints used in this study. 
This work is supported by the "ADI 2021" project funded by the IDEX Paris ANR-11-IDEX-0003-02, and by FAPESP, process numbers 2019/14893-3, 2021/02027-0 and 2021/01089-1.
AS was partially supported by the grant “DS4ASTRO:
Data Science methods for Multi-Messenger Astrophysics
$\&$ Multi-Survey Cosmology”, in the framework of the
PRO3 ‘Programma Congiunto’ (DM n. 289/2021) of
the Italian Ministry for University and Research. AS also acknowledges funding from Next Generation EU, in the context of the National Recovery and Resilience
Plan, Investment PE1 – Project FAIR “Future Artificial
Intelligence Research”. This resource was co-financed
by the Next Generation EU [DM 1555 del 11.10.22]. MJW is supported by the Australian Research Council grants CE200100008 and DP220100007. IR further acknowledges Aion Viana for fruitful discussions.

\bibliographystyle{JHEP}
\typeout{}
\bibliography{bibl}

\end{document}